Chapter 13

# Employing AI and ML for Data Analytics on Key Indicators: Enhancing Smart City Urban Services and Dashboard-Driven Leadership and Decision-Making

*Md Aminul Islam and Md Abu Sufian*

## Abstract

This research navigates the confluence of data analytics, machine learning, and artificial intelligence to revolutionize the management of urban services in smart cities. The study thoroughly investigated with advanced tools to scrutinize key performance indicators integral to the functioning of smart cities, thereby enhancing leadership and decision-making strategies. Our work involves the implementation of various machine learning models such as Logistic Regression, Support Vector Machine, Decision Tree, Naive Bayes, and Artificial Neural Networks (ANN), to the data. Notably, the Support Vector Machine and Bernoulli Naive Bayes models exhibit robust performance with an accuracy rate of 70% precision score. In particular, the study underscores the employment of an ANN model on our existing dataset, optimized using the Adam optimizer. Although the model yields an overall accuracy of 61% and a precision score of 58%, implying correct predictions for the positive class 58% of the time, a comprehensive performance assessment using the Area Under the Receiver Operating Characteristic Curve (AUC-ROC) metrics was necessary. This evaluation results in a score of 0.475 at a threshold of 0.5, indicating that there's room for model enhancement. These models and their performance metrics serve as a key cog in our data analytics pipeline, providing decision-makers and city leaders with actionable insights that can steer urban service management decisions. Through real-time data availability and intuitive visualization dashboards, these leaders can promptly comprehend the current state of their services, pinpoint areas requiring improvement, and make informed decisions to bolster these services. This research illuminates the potential for data analytics, machine learning, and AI to significantly upgrade urban service management in smart cities, fostering sustainable and livable communities. Moreover, our findings contribute valuable knowledge to other cities aiming to adopt similar strategies, thus aiding the continued development of smart cities globally.

## Introduction

Countries across the globe are experiencing rapid urbanisation, which puts tremendous pressure on government officials, real estate developers and businesses. In addition, urban populations are growing at an unprecedented, alarming rate, so they must manage them sustainably and efficiently. Innovative technologies in the form of smart city solutions are the best way to deal with such a situation. These solutions empower government businesses through geospatial and municipal dashboarding and construction companies to improve service efficiencies, create safer neighbourhoods, reduce carbon emissions, save the environment and enhance the overall living standard of citizens (Dameri, 2017). Managing data and making decisions is, however, a challenge for the administration. Having a business intelligence solution to handle such difficulties is quite important. Dashboards are perfect for dealing with this problem. The figure below (Fig. 13.1) represents the smart city dashboard of Lyon, Munich and Vienna (Morishita-Steffen et al., 2021).

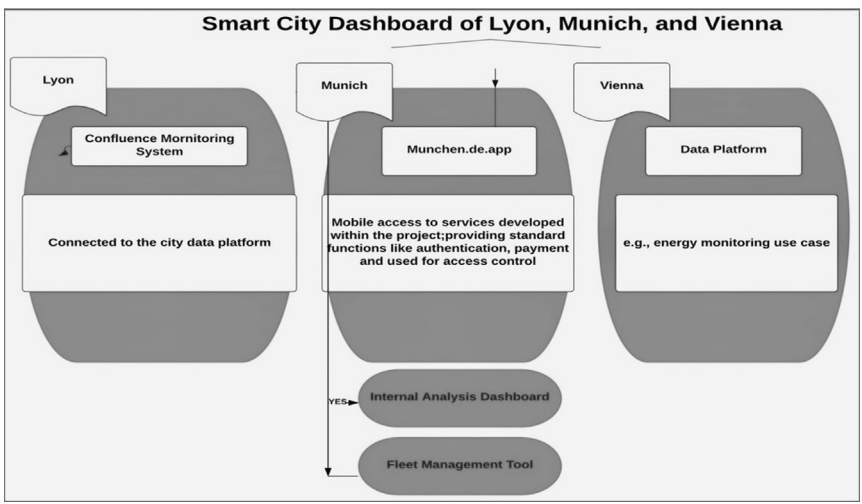

Fig. 13.1.   The Smart City Dashboard of Lyon, Munich, and Vienna.
*Source:* Grand Lyon, 2020.

The visual analytics dashboards that use dynamic images can reveal Data City. Urban systems and infrastructure, as well as a city's society, economy, environment and population, can all be used to characterise it. As new data become available, these graphs serve as the basis for a display that may be updated or changed. For example, a variety of data visualisation formats can be selected, filtered, queried, zoomed in/out, planned, overlaid or visualised simultaneously in several ways (see Fig. 13.1). It is possible to consolidate and arrange the vital data on a single screen under certain circumstances to provide an overview of information (Few, 2006, p. 34), dashboards offer essential information in one view, just like car dashboards and aeroplane dashboards (Dubriwny & Rivards, 2004; Gray, O'Brien, & Hugel, 2016). Users can access a multitude of interrelated dashboards within a single system and perform summary-to-detail explorations for using analytical dashboards. The hierarchical institution facilitates navigation and summary-to-detail exploration. The use of dashboards is growing in mayor's offices, public buildings and specialised websites.

### Importance of Smart City Solution

Analytics services and connected devices have played a vital role in providing complete city information on a centralised platform. A sustainable urban dashboard solution gives business owners access to a tremendous amount of data in the form of charts, diagrams and visuals that cover the entire city when data analytics manage. By analysing police in high-risk areas and identifying trends, interests, concerns and demands, easier-to-use programmers are effective. Decision-making gets strong and more effective as a result.

To achieve digital equity is crucial to offer affordable gadgets and high-speed internet. Locals can obtain dependable internet service through the city's public Wi-Fi hotspots. It is possible to create a plan regarding skills training, affordable devices and providing access to internet services that are affordable using smart city technology.

It is often necessary to invest a lot of money in the maintenance of ageing roads, monuments, bridges and buildings. Innovative technologies are useful for identifying the areas which need repair before infrastructure fails using a detailed report. In bridges and buildings, smart sensors can provide data showing structural changes, such as cracks and tilts. Automatically push notifications are sent to the concerned department for inspection and maintenance (Dameri, 2017).

### Cities Parameters in the Era of Four Industrial Revolutions (4IR)

New technology has been a defining feature of economic and social revolutions and a new way of seeing the world (Dameri, 2017). The 4IRs can be divided into several categories. Products and services that are digitally connected, smart cities and factories, as well as increasing automation at home and work are amongst them. The speed, scope and system impact of progress in this field are distinctive. Since we interact with technology and work in a variety of ways, the fourth IR can make a profound impact on almost every aspect of our lives. With advancements in computing power, efficiency, and decision-making capabilities, the fourth IR is ushering in a digital era.

### First IR

In the first IR, the production of mechanical goods was mechanised using water and steam power. Agriculture dominated the world economy in the seventeenth and eighteenth centuries (see Fig. 13.2) (Kitchin, 2014). Farming dominated everyday life during an agrarian era. Homemade products were commonly made by people using their hands or tools they made themselves. However, because of steam power, these semi-nomadic cultures paved the path for urbanisation. With the advent of steam power and machine tools, the world began to rely more on steamships and railroads for transportation. In the first IR, machines were introduced into the handcrafted textile industry, and mining technologies such as iron and coal extraction were advanced. A skilled middle class was created because of industrialisation. Economies grew along with cities and industries at a faster rate than ever before.

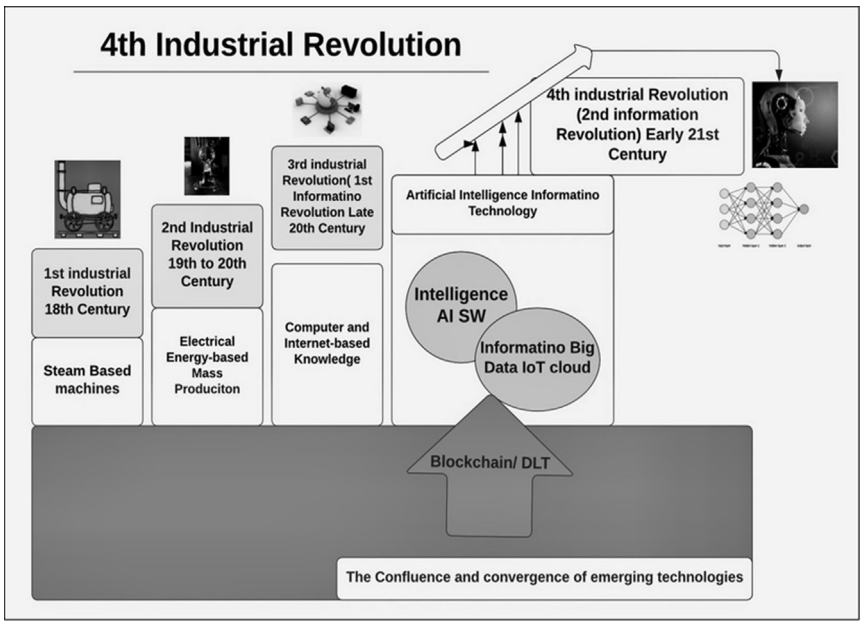

Fig. 13.2.    Fourth Industrial Revolution. *Source:* Adapted from Marsal-Llacuna (2020).

**Second IR**

The second IR occurred after the first. Numerous power sources, including electricity, oil and gas, have been found. In this revolution, mass production was solely obsessed with electric power (see Fig. 13.3). It was the period of gasoline engines, aeroplanes, chemical stimulants, and other inventions were developed (Marsal-Llacuna, 2020). During the second IR, chemical fertilizers increased agricultural production, while coal and iron helped to advance oil and gas mining. With the advent of synthetics, the textile industry progressed within the new era. Several telecommunication technologies were invented, including the telegraph and the telephone. The invention of electric lighting, radios and telephones helped people live and communicate more efficiently as the population grew. Assemblies' lines in the automobile industry are amongst the greatest inventions of the second IR.

**Third IR**

As a result of the introduction of newer sources of power, such as nuclear power, technological advances in computers and the rapid development of electronics, the third IR began in 1969 (Marsal-Llacuna, 2020). During the third IR, advances in electronics, information technology and automation ushered in a modern world with several characteristics. The third IR produced several signs of progress, including the technology of communication through mobile, the wireless internet, emails and the development of medical technologies, biotechnology and pharmaceuticals. Digital technology has altered commonplace analogue electrical and mechanical components, and global communications and energy were particularly affected. With the advent of electronic and information technology, production became automated, and supply chains became global.

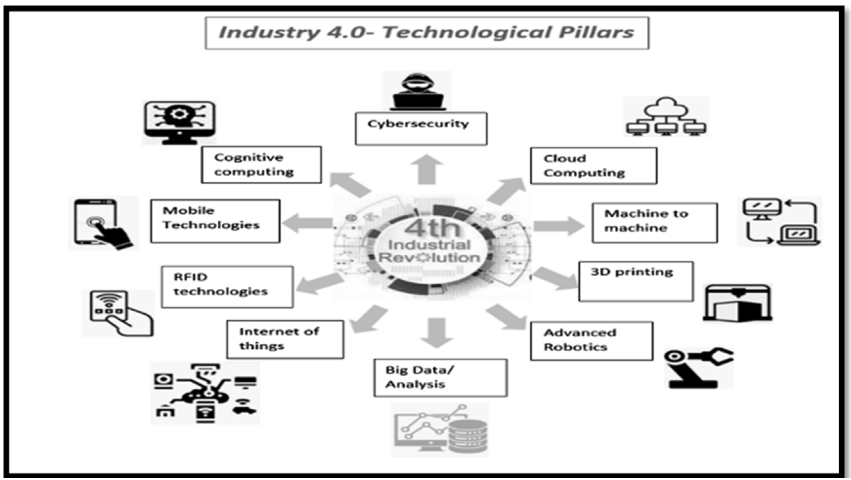

Fig. 13.3.    Digital Fabrication Technologies. *Source:* Marsal-Llacuna (2020).

Information and communication technology (ICT) advancements have transformed rural civilisations into urbanised communities. The human empire has been profoundly affected by information computing technology in every.

The manufacturing sector, open flows of goods, e-business and globalisation, among others. This era was marked by an increase in the value of intellectual property over products and properties. Industrialisation during the third IR, globalisation and the free-market economy grew. Citizens have access to the world at their fingertips. It was during the third IR that the innovations of the two preceding revolts reached maturity.

**Fourth IR**

Physical, digital and biological boundaries are blurring in the fourth IR due to the convergence and fusion of technologies. Internet-based computing offers decentralised and distributed computing in contrast to the third IR's centralised models. A device can store and process an unlimited amount of information because of unprecedented processing power (Marsal-Llacuna, 2020). In the social sphere, there is a paradigm shift taking place. In the next 30 years, our world will become more intelligent, with things sensing and connecting. This will be achieved using leading ICT technologies. The world's astute devices are its sensing organs, its networks are its connections, and its clouds are its digital brains (See Fig. 13.4).

Technologies for digital fabrication are constantly interacting with the world around them. Cloud, network infrastructure (pipes) and devices must work together to achieve industry digitalisation. Internet of Things (IoT), 5G and big data make up the new ICT that combines these foundational, enabler and digital transformation accelerators (See Fig. 13.2).

## The Impacts of Dashboard Smart City

### *The Rise of the Data Dashboards*

A dashboard provides communities, businesses and citizens with an overview of urban development in depth and over time (James, Das, Jalosinska, & Smith, 2020). The usage of a cost-effective data analysis tool is also possible. Smart-city performance dashboards are now a crucial component. For example, in Dublin, residents and decision-makers use and participate in this feature for free. Commuter satisfaction with public transport is used in real-time in Singapore, which is often cited as an example of a smart city.

### *Actions and Transformation of Smart City Dashboard*

Often, modern companies are overloaded with vast amounts of data created by outdated industrial business processes. Data collection and production can be made simpler with the use of modern technology (Jasim, Alrikabi, & Rikabi, 2021). Consequently, we can experience what is known as a data binge, implying that the data is not thoroughly analysed. Converting data into information forthwith is difficult in larger data pools. The shorter product and service life cycles today make this a significant consideration. It is vital to analyse meaningful data and convert it into knowledge, information and ultimately action timely enough to influence an organisation positively. By creating a conceptual framework for turning big data into actionable knowledge, the author provides practical recommendations for agile companies in the twenty-first century. Additionally, this model provides tools for analysing big data to improve decision-making for industry practitioners, as well as a foundation for scholarly research. When evaluating the opportunities associated with accumulating, interacting with, and storing large amounts of data, we should take the resources into account in addition to the economic opportunities needed to assess the risks, such as identifying viruses, resolving them, and setting up internal controls.

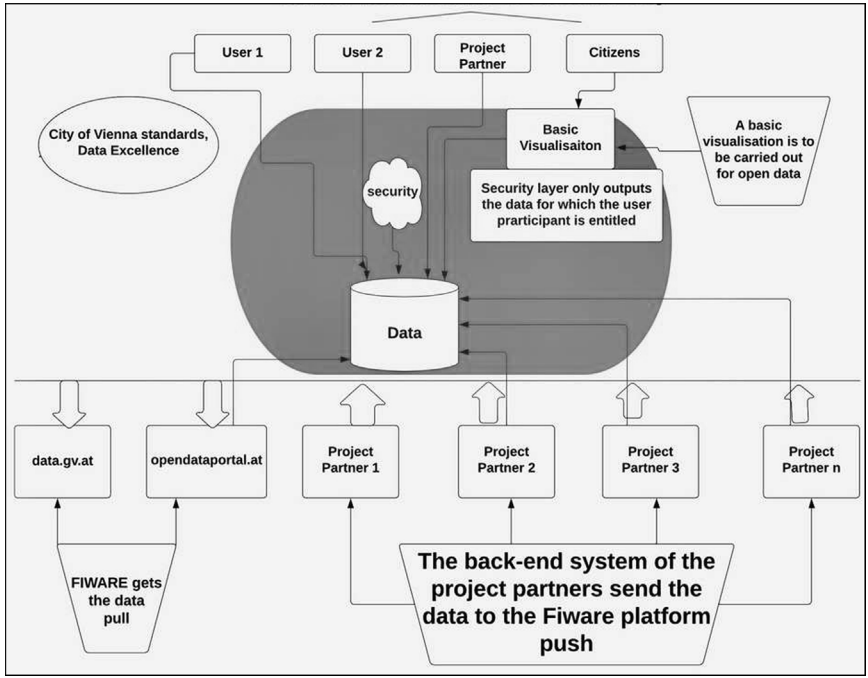

Fig. 13.4.    Smart Dashboard of Vienna City. *Source:* (Smarter Together, 2019).

### Parameters of Good City Dashboard

Data about a city's jurisdiction has long been generated and analysed by city administrations to understand patterns and trends (Marsal-Llacuna, 2020).

However, the data has been widely spread within the organisations from which they originated, and they have been closed. Urban data has been gathered into open data repositories in the transition to open government to make it accessible to everyone through open data repositories (Kitchin, 2014). The availability of urban data is increasing, but skills and literacy are needed for handling, processing, analysing, and visualising them. As a solution, city dashboards have been developed that visualise these data to aid understanding. A city dashboard's purpose is to build trust between public institutions and their citizens (Lněnička & Máchová, 2015). City dashboards are increasingly in demand among a broad range of users, including analysts, policymakers, and politicians.

A good dashboard answers the following questions.

- What is the epistemology behind city dashboards and how are they used for gaining insights and value?
- What is the scope of city dashboards, and how accessible are they?
- How much can we rely on the truth and veracity of city dashboards?
- Is the city dashboard user-friendly and easy to understand?
- In what ways do city dashboards make sense and what are their uses and benefits?
- What ethical principles should we follow to ensure dashboards do no harm and are used responsibly?

## Data Platforms for Smart City Governance

In this section, we discuss how to govern urban data smartly. Scientists, urbanists, policymakers, companies and stakeholders have studied this topic for decades. Using big data and infrastructure, cities like Vienna are becoming nicer. A witty city attempts to mitigate and prevent the challenges presented by 2 billion people moving into urban areas by utilising intelligent technologies and data-driven contextual governance models. Therefore, urban data collection and analysis have become an important task. Using artificial intelligence (AI), municipalities and stakeholders can analyse big data. Policymakers can draw inspiration from many examples of smart cities around the world when it comes to governance and urban planning in smart cities (see Fig. 13.4).

A smart city platform enables multiple resources and categories of data that need to be integrated for access and use by various stakeholders. Increasing data volume, velocity and variety (i.e. big data) are making platforms increasingly critical to transforming disparate data into actionable insights and information. In the same way, smart cities are clarified differently, and platforms are also defined differently. Technically, a platform can be defined as a set of stable components that constrain the relationships amongst other elements so that variety and evolvability can be bolstered in a system. Platforms, such as smart cities, can be defined as 'software-based products or services that act as the basis for other goods or services'. In this definition, the supply chain is viewed as a separate entity from the core functionality that adds value (Fig.13.15).

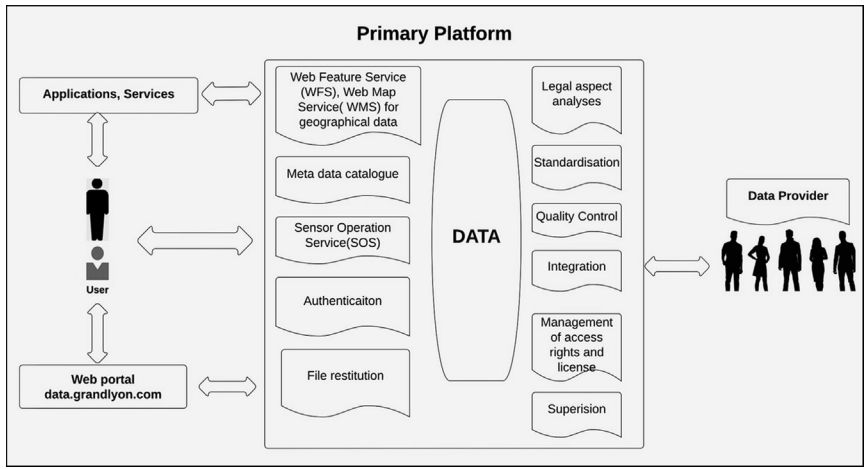

Fig. 13.5.    Primary Platform.

Since 2013, 99% of the data on the Grand Lyon Data Platform has been publicly available. Secure semi-private and private data are protected using a robust authentication mechanism based on standards for Open-Source software (see Fig. 13.7). The smart data platform securely stores personal data.

Confluence is a monitoring system that collects and visualises energy data developed by the Metropole de Lyon in collaboration with Grand Lyon (the Greater Lyon Region).

### Politics Behind City Dashboard

In smart urban areas, city dashboards provide a visual approach to display metrics and progress towards municipal goals. To the uninitiated, city dashboards might seem like the kind of pretty pictures the city posts online as 'news' from time to time. In terms of politics, though, they are beyond that (Jing, Du, Li, & Liu, 2019).

Community, as well as municipal dashboards, provide a structure for governments to update residents and other stakeholders on the development of strategic plans, as has previously been highlighted. Both quantitative findings and qualitative justifications are provided, striking a good balance between raw data and meaningful context. These dashboards are used by cities to showcase their progress, assets, and flaws in an open and transparent manner. Community dashboards are also used by several cities to establish themselves as pioneers by exhibiting their innovative policies and methods. Politically speaking, a government city dashboard has various political features from the standpoint of both residents and local governments, including.

- The practice of holding a government responsible for meeting the needs of its people. It is the duty of a city to better the lives of its citizens and to spread the word about the positive changes it has made. Dashboards are a simple but effective tool for governments to demonstrate that they are listening to the demands of the community and are committed to implementing the changes that have been demanding.
- The process of convincing people that their government is concentrating on the correct things. Municipalities may use city dashboards to highlight accomplishments and outline plans for addressing issues. Citizens will be able to see what the city is keeping tabs on and analysing, as well as the results thus far. This may assist in shedding light on the rationale and context for various municipal activities.
- Keeping the public informed of the government's plans and initiatives. For instance, a citizen may follow an important initiative to show the percentage accomplished each quarter. This helps governments be more transparent with the people they serve, and it maintains lines of contact with those people open (Wiig & Wyly, 2016).

Taking all of this into account, some of the most visible political parts of strategic planning in municipal governments are the dashboards. Based on Jessee (2019), Rossi (2016), Balletto, Borruso, and Donato (2018), Picon (2018), and Kitchin, Maalsen, and McArdle (2016), the following points are elucidated as the government and political perspective of city dashboards.

### Politics Behind Determining Priorities

Prioritisation is a difficult task. The governance of a city is large and costly for a simple reason. Every community must prioritise a wide range of issues and decide where its resources are best allocated. That something is given precedence over everything else if it is designated as a priority by a city is, of course, the very point of strategic planning. Making strategic judgements means potentially alienating certain groups.

### Politics Behind the Solving of the Challenges

For dashboards to be useful, users need to have clear goals in mind. If people are aware of what they are looking for, a city dashboard may provide them with the information they need to see the outcomes and establish priorities. Nevertheless, citizens may hardly expect to get all the responsible citizens to need from a dashboard. Think about the warning lights in people's cars that make people want to pull over so people can consult the handbook. This manual will advise anybody when to call a towing company and when to keep driving, but it will not teach people how to fix the problem itself. Moreover, the same is true for dashboards. They may not provide explicit instructions, but they give individuals the information they need to solve problems on their own.

*Are There Some Critical Issues to Be Addressed in Every Dashboard?*

A strategic plan's results should inform the content of a company's dashboard like.

(1) Public safety: Any municipal dashboard worth its salt should have at least one item related to public safety as seen by its citizens. For instance, two indicators may compare crime and clearance rates to provide information on both the former and the latter.

(2) Economic development: Likewise, as a citizen, people probably want to know how many businesses are in operation in people's communities and whether they are making a profit or a loss.

High-level metrics will reveal how well the municipality is carrying out the objective while allowing consumers to drill down for more measured insights is a necessary feature of any scorecard. A metric is meaningless without some indication of the municipality's evaluation process. For instance, residents can be curious about the correlation between the 'heart attack survival rate' and response times for emergency services.

### How a City Dashboard Becomes Adversely Political

It has been argued that a person needs both moral courage and political resolve to carry off a government dashboard inside a political atmosphere. The best-case scenario, in citizens' view, is when municipal council members and mayors alike support the use of dashboards to improve decision-making and openness. They are cooperative rather than combative because they realise that a dashboard will aid in the management of local data and the narration of a tale. Without the support of upper management, implementing a dashboard may be political with controversial. A dashboard has never been anything but honest, and some politicians are particularly worried about 'bad news.' It is evident that people vote for politicians because they have a specific plan. That is not a terrible thing, but it is important to keep in mind that individuals are only following a procession that has already begun, which might lead to polarisation.

### Reasons Behind a City Dashboard Becoming Unsuccessful From the Government's End

(1) A common problem with dashboards is that they overwhelm users with too much data. Citizens believe that five to six objectives with one to four measures per objective is enough.

(2) The personnel responsible for inputting information into the dashboard cannot do their jobs properly if they do not have access to the necessary information. I have seen dashboards made by local governments where other data are hidden or otherwise inaccessible. Do owners of the measures know how to go about it if they are sent to them every 3 months in PDF format? To what extent will they have access to data that might affect their projections?

(3) One of the elements of a successful dashboard is that users can easily compare data. One's own dashboard data may be compared to that of the whole country, the state, or even simply neighbouring communities.

When looking at this from a political perspective, can these problems be avoided? The answer is yes, and having convenient access to the data in a setting that enables individuals to play with the metrics and acquire extra information, such as reporting software, may help avert these problems. Accountability measures taken by the government are also crucial. The results of people's actions may be beyond their direct control, but they may surely be influenced. If a metric is considerably off goal, for instance, one can investigate what services might be offered or what tactics could be used to improve the reading. The plan and data have finally come together (Balletto et al., 2018; Jessee, 2019; Picon, 2018; Rossi, 2016).

### Benchmarks and Indicators of City Dashboard

Indicators on city dashboards are repeatable quantitative measurements that may be monitored on a regular basis to reveal trends and patterns across time. Rather than being created and utilised in isolation, they are often a component of a larger set of interrelated measures that serve to both verify trends and provide a complete picture of the city. Indicators are seen as a gauge of the development and performance of a city's numerous elements and characteristics. Dashboards often use graphs and maps to highlight these tendencies. They may also be used in different sorts of models to simulate and forecast what would happen under certain conditions and to imagine what the city might look like in its future situations (Kitchin et al., 2016).

The indicators on a city dashboard fall into two broad indicators: social and economic. Regular data collection of urban social indicators for a city dashboard occurs in a representative sample of cities throughout the globe to report on local advancements in the 20 priority areas of the Habitat Agenda. Information for the dashboard is gathered from a variety of sources, including urban observatories at the regional and national levels. Urban social indicators, on the other hand, are a relatively new method of tracking economic and social progress. Although several researchers have used this method to analyse the progress made in established nations, many more have yet to do so (Batty, 2015).

In contrast, a city's economic Indicators may be found on its dashboard and provide valuable insight into the state of the city's economy. The goal of this dashboard is to offer a quick snapshot of important economic indicators such as major sources of income, commercial occupancy, multi-family domestic rates, sales tax data, and workforce and business statistics. The following aspects are gathered based on Young, Kitchin, and Naji (2022), Kitchin et al. (2016), Kitchin and McArdle (2017), Kitchin, Lauriault, and McArdle (2015), Stehle and Kitchin (2020), McArdle and Kitchin (2016), and Mannaro, Baralla, and Garau (2017).

### City Revenue Indicator

The upkeep of municipal services is funded through local taxation and fees. The safety and prosperity of local communities depend on a variety of essential services, such as law enforcement, emergency medical care, building maintenance, and information resources (such as libraries). People often believe that their local property taxes cover the cost of most public services (see Fig. 13.8). This section's goal is to catalogue and categorise the many income streams that support municipal operations. Policies that promote a healthy income stream may be sustained if their key drivers are clearly articulated (Fig. 13.6).

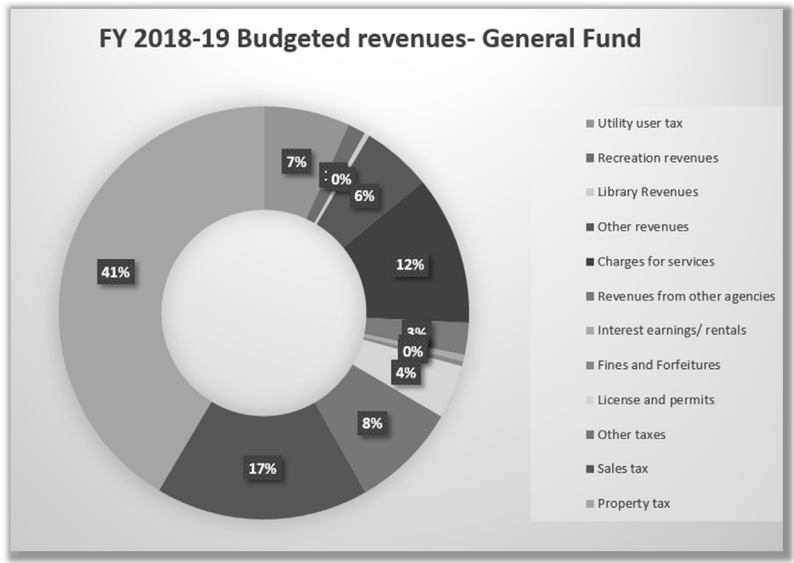

Fig. 13.6.    Example of City Revenue Data.
*Source:* (redwoodcity.org., 2019)

### Sales Tax Indicator

The government imposes a consumption tax on the sale of products and services via a sales tax. The typical retail sales tax is collected by the store and then sent to the government. For sales tax purposes, a company must comply with the rules of each state and locality in which it does business. This 'nexus' might take the form of a physical presence, an employee, an associate, or any other connection to the jurisdiction. Sales taxes, in their traditional retail form, are applied only to the end purchaser. There is much paperwork involved in determining who should pay sales tax since most manufactured goods in modern economies pass through many stages of manufacturing that are each often controlled by a different company. Consider the sheep farmer who sells wool to a knitting mill. The yarn manufacturer must prove to the government that it is not the final user by obtaining a resale certificate to avoid paying sales tax. A resale certificate is required for the yarn manufacturer to sell its product to a garment manufacturer. The manufacturer then sells the woolly socks to a retailer, who must include the sales tax at the final price.

### Employment and Business Indicator

Name, employer, job, location, number of hours, wage or payroll facts, employment figures, and status are all examples of the types of information that may be found in employment data. The term 'business data' refers to all information gathered about a business and its activities (see Fig. 13.7). In this context, 'data' refers to all numerical or numerically based information, such as statistics, raw analytical data, user feedback data, sales figures, and so on. When running a business, it is common to practice conducting more research as feasible to understand better and meet the requirements of customers and employees. Data collection in business may include surveys of clients, statistical analysis, or even just careful observations.

The city dashboard uses benchmarks, which are fixed markers used to determine a location's precise altitude. Surveyors use them as reference points or for measuring site elevations. Urban benchmarking for a city dashboard essentially boils down to a comparison of indicators defining a specific territorial unit (such as a city or metropolitan region) with indicators characterising other units with comparable characteristics. This allows for an accurate assessment of an individual unit as the degree of growth relative to a predetermined standard. Local governments may benefit from urban benchmarking since it helps them conduct evidence-based policy by identifying the most pressing opportunities and problems in a region in connection to established strategic goals. There is a chance to fortify social engagement mechanisms for both the selection of indicators and the distribution of diagnostic findings (see Fig. 13.8).

From the vantage point of a city dashboard, urban benchmarking shines as a tool for the comparative assessment of outcomes in the measurement of complex interactions for which there is no one overarching metric of success. Data acquired in this way is more than simply a bunch of 'meaningless' statistics; it sheds light on how these events compare to others of their kind. The need to measure oneself against others immediately around oneself is innate, and this comparative approach satisfies that need. Learning and adapting to real situations may get off to a solid start with an accurate diagnostic of the current situation (position, outcomes gained, etc.). In addition, benchmarking one's results against

# Workforce Characteristics

### Daily commute inflow and outflow

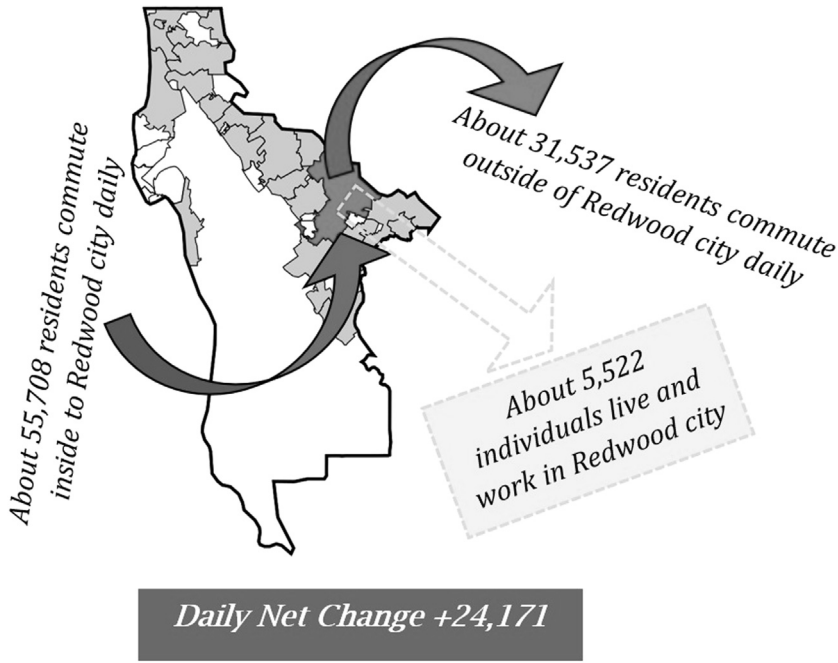

About 55,708 residents commute inside to Redwood city daily

About 31,537 residents commute outside of Redwood city daily

About 5,522 individuals live and work in Redwood city

**Daily Net Change +24,171**

Fig. 13.7.    Employment/Workforce Data Example for California (In-Flow and Outflow Commute Data, for Example). *Source:* US (United States) Census Bureau (2021).

those of others has long been seen as a great approach to developing one's own skills and identifying areas where improvement is needed.

### Importance of Benchmarks and Indicators of City Dashboards

The indicators and benchmarks on a city dashboard may be put to a variety of different uses.

- Regularly refreshed, comprehensible, and quantifiable indicators of the City's present activities in areas like community security or economic growth.
- Past performance indicators are measured, with trends and projections for future improvement or maintenance of the status quo as appropriate.

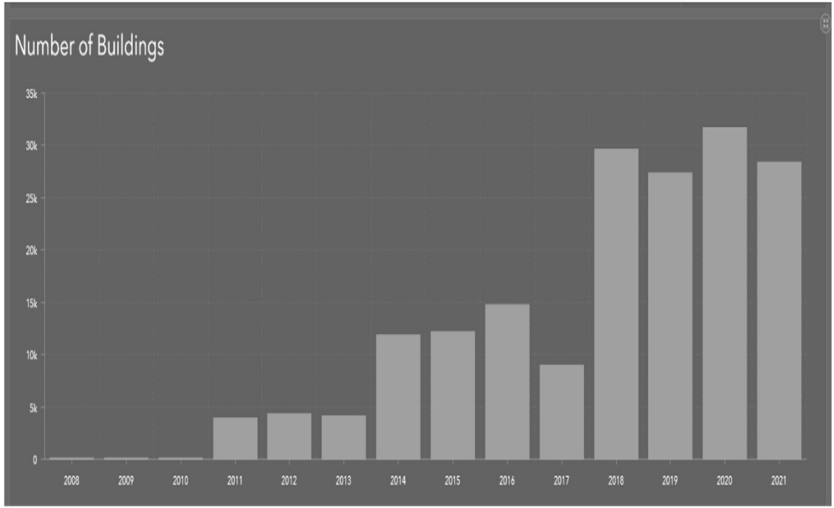

Fig. 13.8.   Example of How Many Buildings Reported Their Benchmarking Data Each Year for a Particular City in a City Dashboard (ArcGIS Dashboards, 2022).

- Indicators of performance that are comparable on a national, state, or even city level are useful benchmarks (Kitchin & McArdle, 2017).

## Tools and Models for City Dashboards

Successful implementation of the current smart city dashboard method necessitates the widespread merging of urban models and tools that previously focused on specific people. The key to comprehending and building a smart city, including a city dashboard in the current day, is strategic planning reference or planning for these domains, as well as synergy between them (McArdle & Kitchin, 2016). The following aspects are integrated as inevitable elements of the models and tools for city dashboards.

### Safe City Models

Physical protection against crime, terrorism, diseases, natural catastrophes, and other threats that may be incorporated into dashboards is a primary concern of the safe city, which also addresses the city's functioning in different circumstances. Models for city safety dashboards include several factors.

- Emergency readiness, damage mitigation, and quick reaction are prioritised in the event of a citywide disaster. Natural disasters, incidents involving hazardous chemicals, pandemics, and other factors may all lead to a state of emergency.

- Create and hone the skills necessary to ensure the smooth running of urban emergency response, commercial, and recovery operations.
- Keeping an eye out for construction/signage, trespassing, and other irregularities that might compromise public safety in an urban setting (Nemani, 2016).

### Digitalisation Model

The need for efficient data gathering is essential to the success of any business, yet the cost of analysing the data is often several times more than the cost of collecting it. In recent years, several municipalities have experienced digital transformation, becoming fully functional digital cities that may be easily included on city dashboards. The 'Digital City' is the smart municipal model's communications and information backbone, and it includes many important ideas that may be included in city dashboards.

- Adapting cutting-edge gadgets into a metropolitan life. Tools for managing traffic flow, trash collection, road maintenance, and garden upkeep are all examples.
- Software for conducting government online; the digitisation of face-to-face processes between citizens and their governments.
- Toolset development is centred on facilitating centralised data integration and data extraction.
- Setting up municipal commands and control hubs for both every day and crisis management.
- Capacity to back up data regularly and restore it quickly in case of an emergency.
- Facilitating a deeper level of communication with locals through online forums and surveys (Granath, 2016).

### The Sustainable City Dashboard Model

Cities may use the Sustainable City Dashboard Model as a tool to monitor their development in the sustainability direction. It provides indications for many different things, including energy usage, travel, trash disposal, and environmental quality. The dashboard may be used to pinpoint areas where cities' sustainability initiatives can be strengthened. In the past, the idea of a 'sustainable city' was discounted as a trendy word, but it is now acknowledged as an integral component of the 'smart city.' The Sustainable City Dashboard Model is an important component of smart cities' utilization of data and technology to enhance the lives of their residents. The work of Dr. F. Bibri, who created a framework for evaluating sustainability in cities, served as the foundation for the dashboard. The following ten sustainability dimensions are included in Bibri's framework: Fig. 13.9. Sustainability Dimensions Framework by Bibri. These factors are used by the Sustainable City Dashboard Model to gauge a city's sustainability. For cities that are devoted to enhancing their sustainability, the dashboard is a useful tool (Bibri, F. 2019).

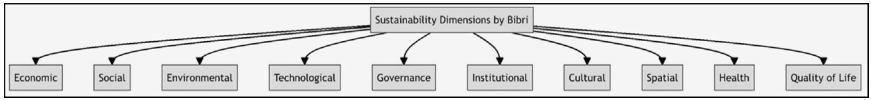

Fig. 13.9.    Sustainability Dimensions Framework by Bibri.

### *The Efficiency Tools and Models for City Dashboards*

Based on several different efficiency plans, the efficient city provides a set of instructions for making the most of the city's current facilities and assets while cutting costs elsewhere. The educational component of these programs should focus on assimilating behavioural adjustments and supporting a healthier approach among the municipality's management, employees, and residents, while the technological component should include behaviour related to the installation of various systems and accessories. The city's ability to deliver high-quality services to its citizens throughout this transition, which can be tracked via city dashboards, must be preserved at all costs. Some models and technologies that may be used to improve urban efficiency are as follows.

• Energy efficiency – Studies show that municipal governments may save anywhere from 10% to 20% of their entire power use just by reducing their electricity costs.
• Organisational memory and a digital inquiry approach to help in dealing with repeated failures.
• The implementation of analytical tools aimed at analysing administrative procedures with the goal of simplifying and streamlining operations (Kourtit & Nijkamp, 2018).

## Demonstration: Machine Learning for City Dashboard

In this section, we used a dataset from Edmonton city of Canada to get a glimpse of real-life analysis using machine learning tools by python code.

### *Feature Selection Method and Extraction*

(1) Environment-related attributes: use of Solar Panels made in Edmonton
(2) Employment: Unemployment rate
(3) Healthy package: impaired driving incidences
(4) More advanced in innovation related any attributes: CMA-labour force growth

(5) Governance related to any dominant attributes: Governance

What kind of Python package should I import into Jupiter Notebook before beginning the machine learning project demonstration is a very crucial stage to arrange all things (see Fig. 13.10).

```
Importing the packages

In [1]:
        import numpy as np               # for multi-dimensional containers
        import pandas as pd              # for DataFrames
        import crayons as cr

        from tabulate import tabulate

        import matplotlib.pyplot as plt
        plt.style.use('classic')
        %matplotlib inline
        import seaborn as sns

        from sklearn import linear_model

        from sklearn.model_selection import train_test_split
        from sklearn.preprocessing import StandardScaler
        from sklearn.neighbors import KNeighborsClassifier
        from sklearn.metrics import classification_report, confusion_matrix

        sns.set(style="white", color_codes=True)
        import warnings # current version of seaborn generates a bunch of warnings that we'll ignore
        warnings.filterwarnings("ignore")
```

Fig. 13.10.   Programme for Importing the Packages.

You will need to import different packages, depending on the specific machine learning project, but the following are some typical packages that are frequently used in machine learning projects.

*NumPy:* This Python package allows for numerical computation. It is frequently used for operations in linear algebra and array manipulation.

*Tabulate:* It is a Python package that is used to create ASCII tables from data in a tabular format. It is useful in machine learning projects to display results in a clear and organised way, and ASCII tables are a way to display data in a tabular format using ASCII characters. ASCII characters are used to create a simple and clean table.

*Pandas:* Data manipulation and analysis are supported by this library. In a tabular format, it is used to load, process and analyse data.

*Matplotlib:* It is used to produce graphs and plots that show the data and outcomes of machine learning algorithms.

*Seaborn:* Visualising data that is based on matplotlib. It offers a sophisticated interface for producing statistical visuals.

*Scikit-learn:* All machine learning projects are supported by this python library. It offers a variety of supervised and unsupervised learning methods, model evaluation, and feature selection.

Most of the project requirements are met with these, which are some of the most frequently used packages for machine learning projects. You may need to import other packages as well, which will depend on the project (Fig. 13.11).

Fig. 13.11.   Loading Data Set.

## Loading Data Set

There are many ways to load a data set in Python, some common ones include

*Using a library:* Python has several libraries that can read various file formats, including CSV, excel files, and Jason; NumPy can read NumPy arrays; and scipy.io can read multiple data formats including mat and arff.

*Reading from a file:* You can achieve this by utilising built-in Python functions like read () and open () to read a file's contents and store them in memory as a string, list or other data structure.

*Downloading from a URL:* You may also use a URL to download data from a distant server. You can issue an HTTP request using the requests library and get the data from the URL. For instance (see Fig. 13.12).

```
import requests
url = "https://raw.githubusercontent.com/yourusername/yourrepo/master/data.csv"
response = requests.get(url)
data = response.text
```

Fig. 13.12.   Importing Request.

*Database access:* If the data are kept in a database, you can connect to it and retrieve the information using tools like sqlalchemy or Pandas. Using sqlalchemy as an illustration (Fig. 13.13).

```
from sqlalchemy import create_engine
engine = create_engine('sqlite:///data.sqlite')
query = SELECT * FROM data
data = pd.read_sql_query(query, engine)
```

Fig. 13.13.   Database Access.

**Information on the Data Set**

```
In [5]: df.info()

        <class 'pandas.core.frame.DataFrame'>
        RangeIndex: 1158 entries, 0 to 1157
        Data columns (total 8 columns):
         #   Column                                        Non-Null Count  Dtype
        ---  ------                                        --------------  -----
         0   UNEMPLOYMENT_RATE                             153 non-null    float64
         1   National_Unemployment_Rate                    153 non-null    float64
         2   Impaired Driving Incidents                    43 non-null     float64
         3   90_RIGHT_ENERGY                               387 non-null    float64
         4   Edmonton CMA - Working Age Population Growth   98 non-null     float64
         5   Edmonton CMA - Labour Force Growth            98 non-null     float64
         6   Edmonton CMA - Employment Growth              98 non-null     float64
         7   governance                                    1158 non-null   int64
        dtypes: float64(7), int64(1)
        memory usage: 72.5 KB
```

Fig. 13.14.   Information the Dataset.

After loading the data set in Python, then we need to know about the dataset. So, we can get information from a dataset in Python by using various data analysis and data manipulation libraries, such as pandas, numpy and scipy. Few ways to get information from a data set in Python (Johnson & Williams, 2019). *Basic Statistics:* you can use libraries like pandas or numpy to get basic statistics about the data, such as mean, median, standard deviation and others. For example, using pandas (Fig. 13.15).

```
import pandas as pd
data = pd.read_csv('data.csv')
mean = data.mean()
median = data.median()
standard_deviation = data.std()
```

Fig. 13.15.   Importing Using Panda.

*To Summaries data:* you can use the describe () method in pandas to get a summary of the data set's distribution. For example, summary = data. describe ()

*Dicing and Slicing:* you can extract specific subsets of the data based on certain conditions using the loc and iloc methods in pandas. For example, data_subset = data.loc[data['column_name'] == value]

The actions depend on the nature of the data that you are using and the questions you are attempting to answer.

## Data Visualisation and Analysis (Based on Data Set Types) and Exploratory Analysis

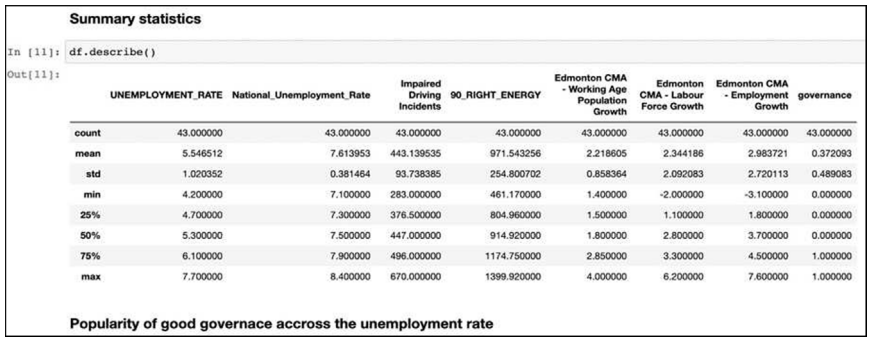

**Summary statistics**

In [11]: `df.describe()`

Out[11]:

| | UNEMPLOYMENT_RATE | National_Unemployment_Rate | Impaired Driving Incidents | 90_RIGHT_ENERGY | Edmonton CMA - Working Age Population Growth | Edmonton CMA - Labour Force Growth | Edmonton CMA - Employment Growth | governance |
|---|---|---|---|---|---|---|---|---|
| count | 43.000000 | 43.000000 | 43.000000 | 43.000000 | 43.000000 | 43.000000 | 43.000000 | 43.000000 |
| mean | 5.546512 | 7.613953 | 443.139535 | 971.543256 | 2.218605 | 2.344186 | 2.983721 | 0.372093 |
| std | 1.020352 | 0.381464 | 93.738385 | 254.800702 | 0.858364 | 2.092083 | 2.720113 | 0.489083 |
| min | 4.200000 | 7.100000 | 283.000000 | 461.170000 | 1.400000 | -2.000000 | -3.100000 | 0.000000 |
| 25% | 4.700000 | 7.300000 | 376.500000 | 804.960000 | 1.500000 | 1.100000 | 1.800000 | 0.000000 |
| 50% | 5.300000 | 7.500000 | 447.000000 | 914.920000 | 1.800000 | 2.800000 | 3.700000 | 0.000000 |
| 75% | 6.100000 | 7.900000 | 496.000000 | 1174.750000 | 2.850000 | 3.300000 | 4.500000 | 1.000000 |
| max | 7.700000 | 8.400000 | 670.000000 | 1399.920000 | 4.000000 | 6.200000 | 7.600000 | 1.000000 |

**Popularity of good governace accross the unemployment rate**

Fig. 13.16.    Data Visualization and Analysis (Based on Dataset Types) And Exploratory Analysis.

With the use of a statistical methodology, we were able to summarise the data from the previous section using Python's code data frame to describe the method. In a dataset where every attribute value is statistically represented (Figs. 13.14, 13.16).

### The Popularity of Good Governance Across the Unemployment Rate
An important driver of economic growth, good governance has a direct impact on a geopolitical region's GDP (Gross Domestic Product), employment rate and inflation rate (see Fig. 13.15). The unemployment rate in Edmonton is the same for people who believe that the governance is being done properly and for people who believe that it is moving in the right direction.

### Assessing the Presence of Extreme Values and Equal Variance
Assessing the presence of extreme values and equal variance are important steps in exploratory data analysis (EDA) for smart city dashboard making. Because they can have a significant impact on the results of statistical tests and models. Here's why:

Extreme Values: Extreme values, usually referred to as outliers, can significantly affect the outcomes of statistical tests and models. For instance, a single extreme number in a dataset can skew the mean and standard deviation, making it challenging to understand the findings. To prevent bias in the results, it's critical to recognise and deal with outliers in a data set.

Equal Variance: Many statistical tests and models assume equal variance, often known as homoscedasticity. Equal variance denotes that the data's variability is constant over the whole range of the independent variable. The test and model findings might not be accurate if the variance is not equal. Uneven variance, for instance, might lead to inaccurate estimations of the coefficients in a linear regression model.

There is no presence of extreme values across the two groups, that is, those that think the government of Edmonton is in a good direction or good governance and those having a contrary opinion (see Fig. 13.17). Generally, there is no variation in the use of solar panels across the divide. This means that both groups conserve the environment.

### Normality Using Histogram

Histograms can be created in Python using various libraries, such as matplotlib and seaborn. It is a bar graph that represents the frequency of each value in a

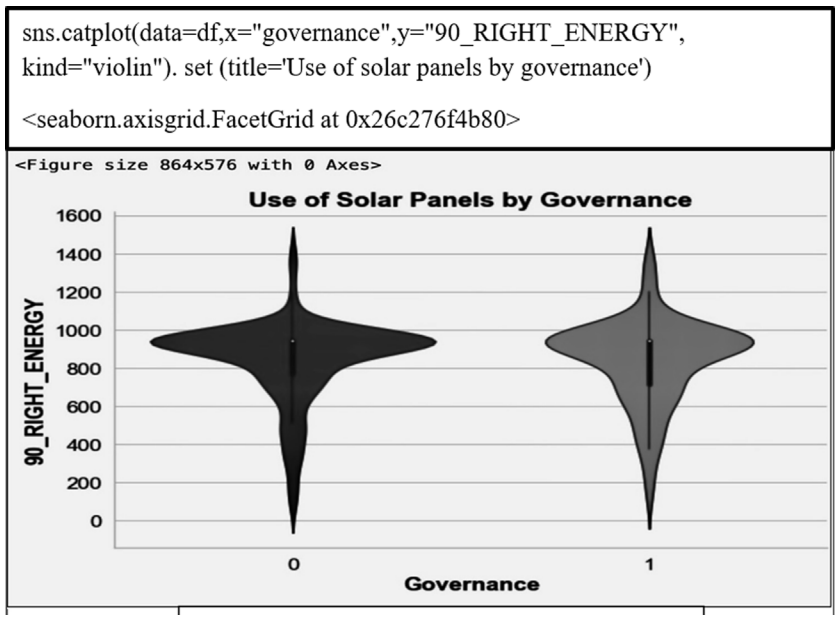

Fig. 13.17.    The Use of Solar Panels by The Government.

dataset. A data set in a Python project, why a histogram is necessary for. There are some reasons which have been mentioned below:

*Understanding the distribution:* A histogram can help you identify patterns, skewness, outliers and other important features of the distribution.
*Identifying outliers:* Outliers are values that lie significantly away from the majority of the data. They can have a significant impact on the results of statistical tests and models. A histogram can help you quickly identify outliers and decide how to deal with them.
*Selecting appropriate statistical models:* Different statistical models are appropriate for different distributions. A histogram can help you determine which model is appropriate for your data by showing you the shape of the distribution.
*Checking assumptions:* Many statistical tests and models make assumptions about the distribution of data. A histogram can help you check these assumptions and determine whether your data are suitable for the test or model you want to use.
*Communicating results:* A histogram is a simple and effective way to communicate the distribution of the data to others. By showing the frequency of each value, a histogram makes it easy to see the patterns and trends in the data.

Avoid having the matplotlib verbose information.

The results show that the assumption of normality is violated across all the variables as the shape of the histograms can be estimated to be bell-shaped (see Figs. 13.19).

### Data Evaluation Is Like Cleansing and Transforming

**Data Pre-processing Method**

Preparing and cleaning the data before feeding it into the model is known as 'data pre-processing', and it is a crucial stage in the machine-learning pipeline. It is essential since the performance of the model is directly correlated with the calibre of the data used to feed it (see Fig. 13.19).

The key procedures in data preparation are as follows.

(1) Identifying and replacing or eliminating missing values from the dataset is the process of handling missing values.
(2) Outliers are extreme values that can affect the performance of the model. To make them more comparable to the other values in the dataset, they can be altered or eliminated.
(3) Feature scaling entails scaling down the features to make them more comparable. This is significant because the scale of the features can affect the performance of various machine learning methods.
(4) Categorical variables can only have a certain number of possible values, hence they must be encoded. Before they can be entered into the model, they must be transformed into numerical values (Fig. 13.18).

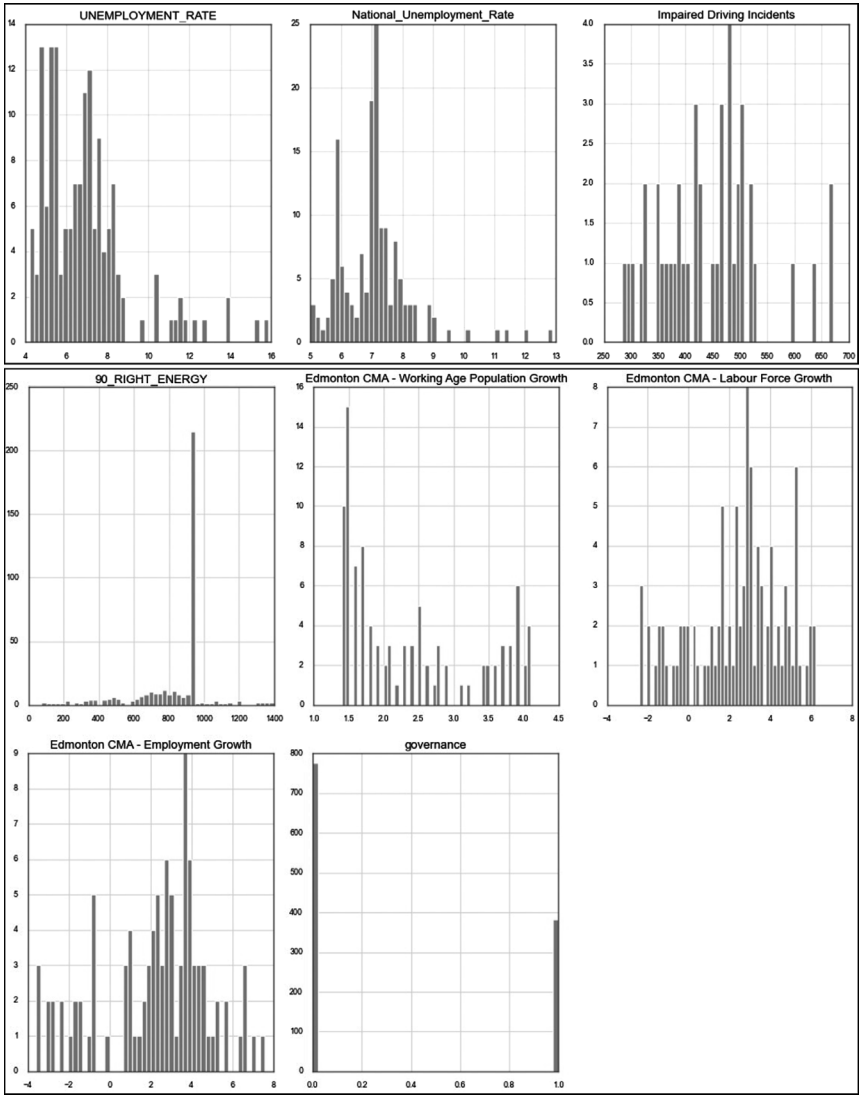

Fig. 13.18.    Normality Using Histogram.

(5) Data division into training and test sets: The data are divided into two sets: a training set for training the model and a testing set for assessing the model's efficacy to ensure that the model is appropriate for its intended use and trained on high-quality data (see Fig. 13.19).

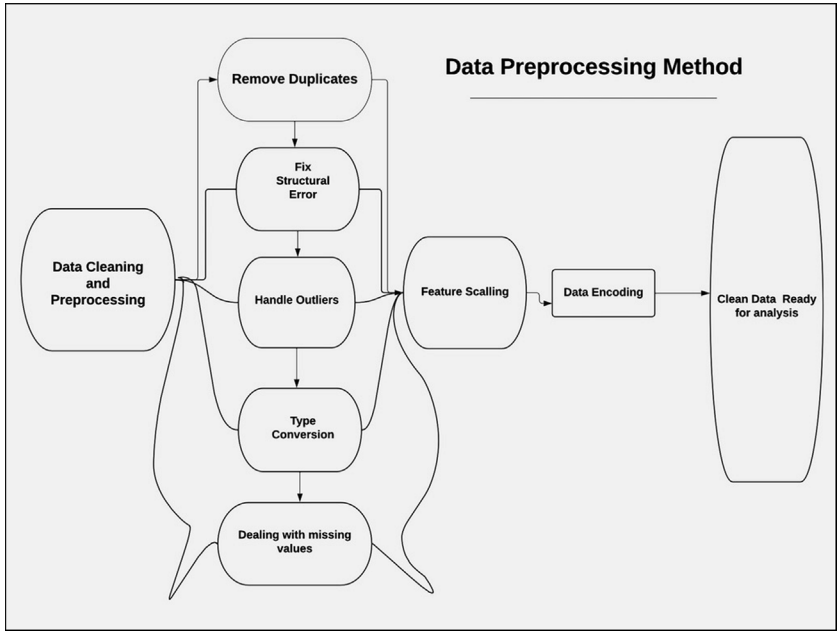

Fig. 13.19.    Data Pre-processing Method (Mallick, S. (2022)).

```
In [60]: df.isnull().sum()

Out[60]: UNEMPLOYMENT_RATE                              1005
         National_Unemployment_Rate                     1005
         Impaired Driving Incidents                     1115
         90_RIGHT_ENERGY                                 771
         Edmonton CMA - Working Age Population Growth    1060
         Edmonton CMA - Labour Force Growth             1060
         Edmonton CMA - Employment Growth               1060
         governance                                        0
         dtype: int64

In [61]: df=df.dropna()

In [62]: df.isnull().sum()

Out[62]: UNEMPLOYMENT_RATE                              0
         National_Unemployment_Rate                    0
         Impaired Driving Incidents                    0
         90_RIGHT_ENERGY                               0
         Edmonton CMA - Working Age Population Growth   0
         Edmonton CMA - Labour Force Growth            0
         Edmonton CMA - Employment Growth              0
         governance                                    0
         dtype: int64
```

Fig. 13.20.    Dealing with Missing Values Field.

The data set had several missing values across all the variables. The missing values were dropped as there was no better alternative that could be deployed, and the sample size was large enough (Fig. 13.20).

**Split the Values Into *x* and *y***

```
In [74]: from sklearn import preprocessing
         from sklearn import utils

         X = df.iloc[:, :-1].values    #   X -> Feature Variables
         y = df.iloc[:, -1].values #   y -> Target
```

Fig. 13.21.    Split the values into x and y.

An important stage in many machine-learning algorithms is dividing the values into $x$ and $y$ before making a model (Fig. 13.21). The independent variables are represented by $x$, and the dependent variables are represented by $y$ (see Fig. 13.22). It enables you to forecast the dependent variable and fit the model to the independent variables.

Cross-validation to assess the model's performance by dividing the data into $x$ and $y$. In cross validation, the data are divided into training and testing sets, with the training set being used to fit the model and the testing set is used to assess how well it performs.

**Splitting the Data Into Train and Test**

Splitting the data set into a training set and a test set is an important step in the machine learning process because it helps you to evaluate the performance of your model. The idea is to divide the data into two parts: a training set that is used to fit the model and a test set that is used to evaluate the performance of the model.

The significance of splitting the data into train and test is that it provides an estimate of how well the model will perform on unseen data. By comparing the predictions of the model with the actual values in the test set, you can estimate the accuracy of the model and determine whether it is overfitting or underfitting the data.

Here the data has been divided into two sets, 70% for the training set and 30% for the testing set. The 70–30 split has been chosen for this accuracy over other splits for the Edmonton data set.

$X$_train, $X$_test, $y$_train, $y$_test = train_test_split ($X$, $y$, test_size = 0.3, random_state = 0).

A data set into training and test sets for dividing, there is no predetermined proportion. The size of the data set, the difficulty of the challenge and the type of data are frequently considered when determining the split %. The typical split ratios are 60/40–80/20, with most of the data being utilised for training and the rest for testing.

For instance, a smaller test set might be adequate if the data set is huge and the issue is straightforward. On the other hand, a bigger test set can be required if the data set is limited or the issue is complex to provide a trustworthy assessment of the model performance.

```
In [67]:  scaler = StandardScaler()
          scaler.fit(X_train)

          X_train = scaler.transform(X_train)
          X_test = scaler.transform(X_test)
```

Fig. 13.22.    Standardize Features.

**Standardise Features by Removing the Mean and Scaling to Unit Variance**
Standardising features by removing the mean and scaling to unit variance is an important pre-processing step in many machine learning algorithms (see Fig. 13.23). By scaling the features to have zero mean and unit variance, the features are centred around zero and have the same order of magnitude. Normalisation and standard features make it easier for machine learning algorithms to find suitable patterns in the data and can progress the accuracy and speed of training. Additionally, standardising can help mitigate the effects of outliers and make the data more interpretable. It is also often required for certain algorithms that assume standardised features, such as principal component analysis (PCA) or linear discriminant analysis (LDA).

*Type of Machine Learning Algorithms*

Since the target variable governance is binary in nature, the suitable machine-learning algorithm is classification.

The act of identifying, comprehending and organising concepts and objects into predetermined categories or 'sub-populations' is the process that is referred to as classification. Machine-learning systems utilise a variety of methods to classify future data sets into categories. These algorithms are trained using data sets that have already been categorised.

In the field of machine learning, classification algorithms make use of input training data to make predictions about the likelihood that subsequent data will fall into one of the established categories. In this situation, we will evaluate the quality of governance in Edmonton based on several additional criteria to determine whether it is good or terrible.

The algorithms that will be used include.

(1) Logistic Regression
(2) Support Vector Machines
(3) Decision Tree
(4) Naive Bayes

In mathematical notation, the logistic regression model can be expressed as:

$$P(Y = 1|X_1, X_2, ..., X_p) = \sigma(\beta_0 + \beta_1 X_1 + \beta_2 X_2 + ... + \beta_p X_p)$$

Fig. 13.23.    Mathmetical Notation of Logistic Regression.

**Logistic Regression**

Logistic regression is a statistically supervised learning algorithm. It is used where the problems are related to binary classification. The goal is to predict a binary outcome, such as whether a city will call a digital city or not. It is suitable for making a smart city dashboard because the goal is to predict a binary outcome, such as whether a specific event will occur or not. In a smart city context, this might involve predicting the likelihood of a traffic jam, crime or other events based on available data.

In mathematical notation, the logistic regression model can be expressed as

$$P(Y = 1|X_1, X_2, \ldots X_P) = \sigma(\beta_0 + \beta_1 X_1 + \beta_2 X_2 + \ldots + \beta_P X_P)$$

where $\beta_0$ constant and $\beta_1, \beta_2, \ldots + \beta_P$ are the regression coefficients that need to be estimated from the training data.

Logistic regression can be trained on a large amount of data quickly and easily which is useful for leadership decision-making in a smart city context. Furthermore, it is suitable for working with diverse and complex data sources, such as those found in a smart city. This can include data from various sensors, social media feeds and other sources. In logistic regression, the input features are linearly combined and transformed using the logistic function to produce a probability value between 0 and 1. This probability value is then the threshold to make the final binary prediction. In machine learning, thresholding is commonly used in classification tasks, where a threshold is used to determine whether a predicted probability or score should be classified as positive or negative. For example, if a classification model predicts the probability of a patient having a disease, a threshold value can be set to determine whether the patient should be classified as having the disease or not.

This supervised machine learning is trained by minimising the cost function that measures the error between the predicted probabilities and the true binary labels. The most common optimisation algorithm used for logistic regression is gradient descent. However, it assumes that the relationship between the input features and the output variable is linear and may not perform well if the relationship is complex or nonlinear.

**Support Vector Machines**
Support vector is a supervised machine (SVM) learning algorithm which is used for different purposes, but this implies after the logistic regression to see how the result gives for smart city dashboard leadership decision-making because it is a powerful and effective machine learning algorithm that can classify data accurately and efficiently. It is particularly used where the data are complex and nonlinear in smart city applications. SVM can help leaders make informed decisions by providing insights and predictions based on large volumes of data collected from various sources in the city (Smith, 2021, 2022). Also, it improves the services and infrastructure and enhances the quality of life for citizens. Support vector machine (SVM) is a machine learning algorithm used for classification and regression analysis by finding a hyperplane that maximally separates the data into different classes. SVM can also be used for regression, where it finds a hyperplane that best fits the data.

In classification, it finds a hyperplane that maximally separates the data into different classes. Given a set of training examples $(X_1, Y_1), (X_2, Y_2), \ldots, (X_n, Y_n)$, where $X_1$ represents the input features and $Y_1$ represents the corresponding output labels, SVM finds the hyperplane that maximises the margin between the two classes. Mathematically, this can be formulated as.

Maximise: $(W, B)$
Subject to: $Y_i(W\text{^}T\ X_i + B) >= 1, i = 1, ..., n$

$$|W| = \text{sqrt}(W\text{^}TW)$$

where $W$ represents the weight vector, $B$ represents the bias term, and $|W|$ is the norm of the weight vector. The hyperplane is defined by the equation $W\text{^}T\ x + b = 0$, where $x$ is a new input example. The sign of the output of this equation determines the predicted class for the input example.

**Decision Tree**
The decision tree algorithm is a supervised learning algorithm that creates a tree by repeatedly partitioning the data into smaller subsets based on the values of the input features and then assigning a label to each subset based on the majority label of the examples in that subset. The resulting tree can then be used to predict the output label for new input data. We used this algorithm to see the outcomes and the way it follows and shows after the logistics and SVM.

It is often used for demonstrating a project's data analytics on key indicators for smart city urban services by machine learning because it is an effective tool for analysing and visualising complex data for classification and regression tasks, and they provide an easy and intuitive way to represent the decision-making process. In the context of smart cities, decision trees can help city leaders to better understand and optimise the use of resources, predict demand for services, identify areas where improvements can be made for non-technical stakeholders for the decision-making process and communicate the results to a wider audience.

**Naive Bayes**

Naive Bayes is a traditional, statistical but potential machine-learning algorithm that is generally used for classification tasks, such as text classification, spam filtering and sentiment analysis. It is based on Bayes' theorem, which is a statistical formula for calculating the probability of an event occurring, given prior knowledge.

Bayes' theorem is a fundamental stage in probability theory, which provides a way to calculate the conditional probability of an event and give some prior knowledge or evidence.

The theorem states that:

$$P(A|B) = P(B|A) * P(A) / P(B)$$

where:

$P(A|B)$ is the probability of event $A$ occurring, given the evidence $B$

$P(B|A)$ is the probability of observing evidence $B$, given that event, $A$ has occurred

$P(A)$ is the prior probability of event $A$ occurring before any evidence is observed

$P(B)$ is the total probability of observing the evidence $B$, across all possible values of $A$

In Naive Bayes, the algorithm assumes that the features (or attributes) of the data are independent of each other, given the class label. This is known as the 'naive' assumption, and it simplifies the calculation of the probability of a particular class label, given the observed features. The probability calculations are based on a training set of labelled examples, where the algorithm learns to estimate the probabilities from the observed frequencies of the features in each class for large datasets, and it can work well even with limited training data. It is an important machine-learning algorithm in smart city decision-making because it can classify and predict outcomes quickly and accurately, even with limited training data. This is particularly valuable in the context of smart cities, where there is often a large volume of data and a need for fast and efficient decision-making. Naive Bayes can be used to classify various types of data, such as traffic flow patterns, energy consumption and air quality, which are critical factors in smart city decision-making. The algorithm can also be used to predict future trends, such as changes in weather patterns, traffic congestion or the demand for certain services and their application to enhance the quality of life for citizens.

**Evaluation Metrics**

We will evaluate the performance of the model using two metrics – Confusion matrix and accuracy.

A confusion matrix, also known as an error matrix, is a special table structure that permits visualisation of the performance of an algorithm, typically a supervised learning one. It is used in the field of machine learning, and more specifically, the problem of statistical classification (in unsupervised learning it is usually called a matching matrix). Whilst each row of the matrix represents the cases that belong to an actual class, each column of the matrix represents the instances that belong to a predicted class or vice versa – the literature refers to both permutations. The origin of the name comes from the fact that it is simple to determine whether the system is mixing two different classifications (i.e. commonly mislabelling one as another). It is a specialised form of the contingency table that has two dimensions (called 'actual' and 'predicted') with identical sets of 'classes' in both dimensions (each combination of dimension and class is a variable in the contingency table).

Accuracy: The term 'accuracy' refers to the degree to which a binary classification test correctly detects or eliminates a condition from consideration. Accuracy is measured statistically. In other words, accuracy refers to the proportion of right predictions (including both true positives and true negatives) made from the total number of cases that were investigated. As such, it evaluates the differences between the pre-test and post-test probabilities. It is frequently referred to as the 'Rand accuracy' or the 'Rand index' to make the semantics of the context more obvious. It is something that the test uses as a parameter. The following formula can be used to quantify the correctness of binary data:

$$\text{Accuracy} = \frac{\{TP + TN\}}{\{TP + TN + FP + FN\}}$$

where $TP$ = True positive; $FP$ = False positive; $TN$ = True negative; $FN$ = False negative.

Ideally, higher accuracy values are indicative of a good model.

**Data Modelling**

The process of representing the data in the form of diagraming data so as to convert data into information and wisdom by applying some logical, conceptual or mathematical or other formal techniques.

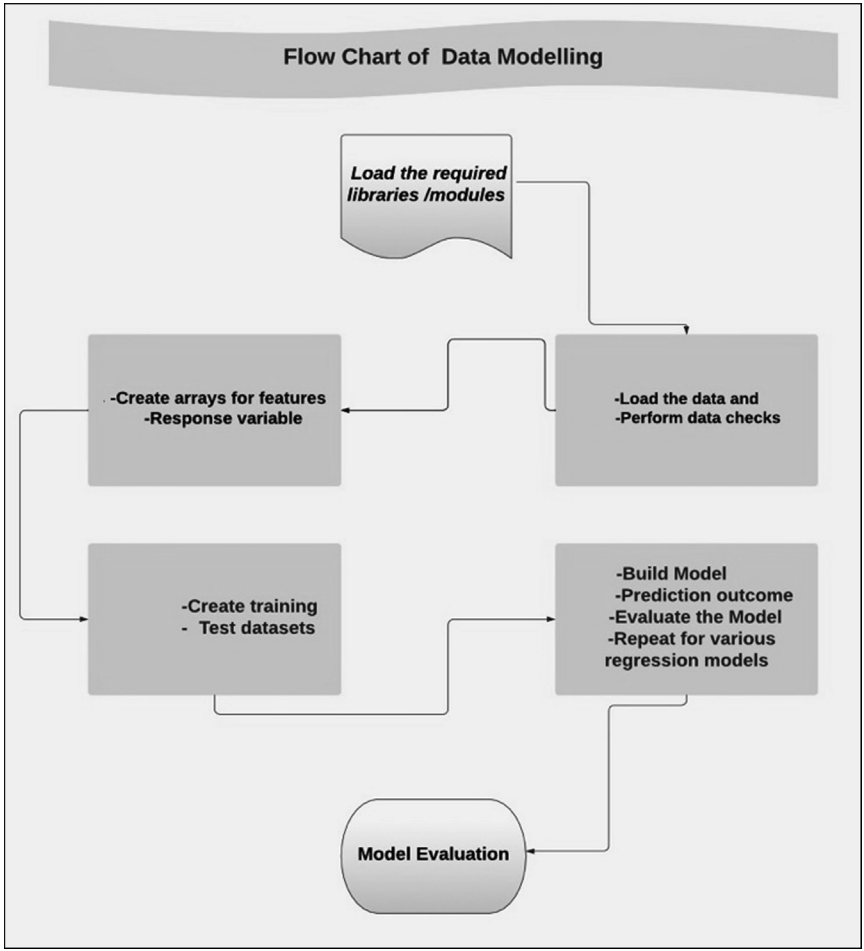

Fig. 13.24.    Data Modelling Stage.

## Building the Models

### Logistics Regression

A *confusion matrix* is a table that is used to evaluate the performance of a machine-learning model. It displays the number of true positives, false positives, true negatives and false negatives that the model has predicted (Fig. 13.25).

In a logistic regression analysis, the confusion matrix helps to determine the accuracy of the model's predictions by comparing the predicted classes with the actual classes of the data. It assists to identify the instances where the model is making correct predictions and where it is making errors, allowing for further refinement of the model.

**Logistics Regression**

```
In [77]:  from sklearn.linear_model import LogisticRegression

          classifier = LogisticRegression()
          classifier.fit(X_train, y_train)

          y_pred = classifier.predict(X_test)

          # Summary of the predictions made by the classifier
          print(classification_report(y_test, y_pred))
          print(confusion_matrix(y_test, y_pred))
          # Accuracy score
          from sklearn.metrics import accuracy_score
          print('accuracy is',accuracy_score(y_pred,y_test))

                        precision    recall  f1-score   support

                     0       0.62      0.56      0.59         9
                     1       0.20      0.25      0.22         4

              accuracy                           0.46        13
             macro avg       0.41      0.40      0.41        13
          weighted avg       0.49      0.46      0.48        13

          [[5 4]
           [3 1]]
          accuracy is 0.46153846153846156
```

Fig. 13.25.    Logistics Regression.

In the confusion matrix, three major subjects are conducted. Precision, recall and $F_1$ score are of the commonly used metrics that are calculated from a confusion matrix to evaluate the performance of a machine-learning model.

*Precision:* It measures the proportion of predicted positives that were true positives,

$$\text{Precision} = \text{True Positives}/(\text{True Positives} + \text{False Positives})$$

*Recall*, also known as sensitivity or true positive rate, is the ratio of true positives to the total number of actual positives in the data. It measures the proportion of actual positives that were correctly identified by the model, and it indicates the model's ability to detect positive instances in the data.

$$\text{Recall} = \text{True Positives}/(\text{True Positives} + \text{False Negatives})$$

*$F_1$ score* depends on precision and recall, and it provides a single measure of the model's overall performance. It balances precision and recall by taking their weighted average, where the emphasis is placed on the lower value. A high $F_1$ score indicates that the model has both good precision and good recall, while a low $F_1$ score indicates that the model is lacking in one or both measures.

$$F_1 \text{ Score} = 2 * ((\text{Precision} * \text{Recall})/(\text{Precision} + \text{Recall}))$$

Where:

True positives (TP) are the number of correctly predicted positive instances.
False positives (FP) are the number of negative instances that were incorrectly predicted as positive.
False negatives (FN) are the number of positive instances that were incorrectly predicted as negative.

A high precision score indicates that a model is making accurate predictions. However, the optimal balance between precision and recall may vary depending on the specific problem being addressed.

A high precision score means that a high proportion of the predicted positive instances are true positives, i.e. the model is avoiding false positives. This is important in situations where false positives can have a significant impact, such as in medical diagnosis or fraud detection.

In general, a high $F_1$ score, which is a combined measure of precision and recall, is used to assess the overall performance of a model. However, the relative importance of precision and recall may depend on the specific application and should be determined based on the cost or impact of different types of errors in the problem being addressed.

If a smart city dashboard for leadership and decision-making model has a high precision score, it would mean that the model is accurately identifying positive instances, i.e. the model is avoiding false positives.

In the context of a smart city dashboard, this could mean that the model is accurately detecting and reporting on events, trends or issues of interest to city leaders, without generating a lot of false alarms. For example, if the model is used to identify traffic congestion in the city, a high precision score would mean that the model is accurately identifying instances of congestion without generating many false reports of congestion, which could help city leaders make informed decisions about traffic management and infrastructure investments.

However, it is important to note that a high precision score alone may not be sufficient to assess the performance of the model, and other metrics such as recall and $F_1$ score should also be considered to evaluate the model's overall effectiveness in achieving the goals of the smart city dashboard.

Above from Fig. 13.22, accuracy shows 0.46 mean 46%. It does not only express performance because a model depends on precision, recall and $F_1$ score as well. So, the meaning of precision score 0.49 means 49%, respectively recall 46% and $F_1$ 48% in a model. These score percentages explain value as I already discussed above respectively.

**Support Vector Machine**

As the above, discussed the SVM algorithm, and I have defined it (Fig. 13.27), why it's used and its significance. From this table accuracy of 0.69 mean 69%. And precision, recall and $F_1$ score respectively in detail.

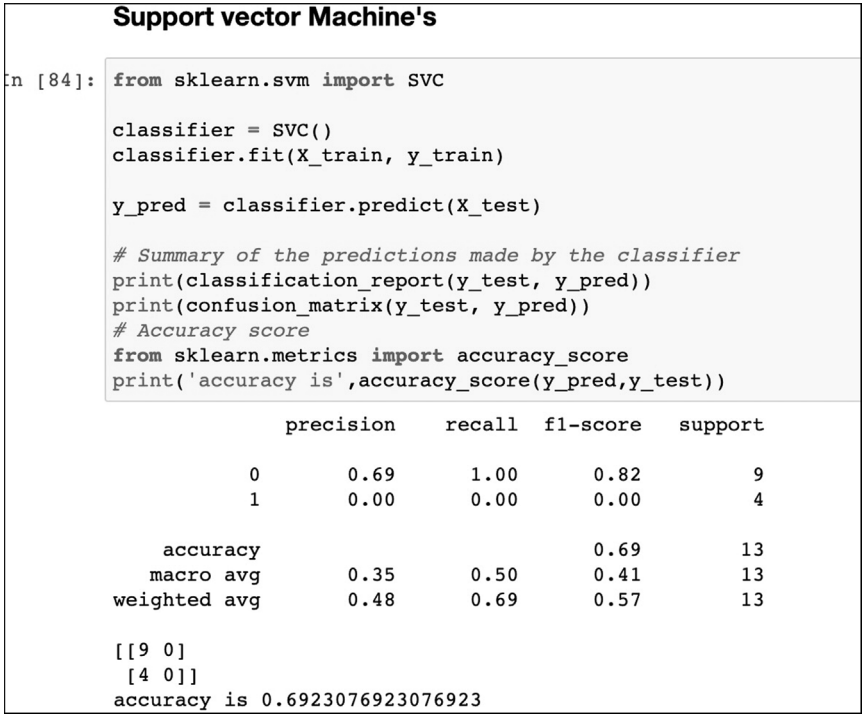

**Support vector Machine's**

```
from sklearn.svm import SVC

classifier = SVC()
classifier.fit(X_train, y_train)

y_pred = classifier.predict(X_test)

# Summary of the predictions made by the classifier
print(classification_report(y_test, y_pred))
print(confusion_matrix(y_test, y_pred))
# Accuracy score
from sklearn.metrics import accuracy_score
print('accuracy is',accuracy_score(y_pred,y_test))
```

```
              precision   recall  f1-score   support

           0       0.69     1.00      0.82         9
           1       0.00     0.00      0.00         4

    accuracy                          0.69        13
   macro avg       0.35     0.50      0.41        13
weighted avg       0.48     0.69      0.57        13

[[9 0]
 [4 0]]
accuracy is 0.6923076923076923
```

Fig. 13.26.    Support Vector Machine Model Result by Performance Metrics.

**Decision Tree**

From this Fig. 13.24 0.53 mean 53% accuracy. Let's start with the formulas for precision, recall and $F_1$ score.

*Precision* = true positives/(true positives + false positives)
*Recall* = true positives/(true positives + false negatives)
*$F_1$ score* = 2 * (precision * recall)/(precision + recall)

Why have we considered weighted average score? Because of considering a weighted average of precision, recall and $F_1$ score can balance the trade-offs between precision and recall and get a more comprehensive evaluation of the model's performance (Fig. 13.26).

We can explain in a mathematical way. Let's denote the weights as $w_1$, $w_2$ and $w_3$, respectively. Then, the weighted average can be calculated as follows (Fig. 13.27):

Weighted average $= (w_1 * \text{precision} + w_2 * \text{recall} + w_3 * F_1 \text{ score})/(w_1 + w_2 + w_3)$.

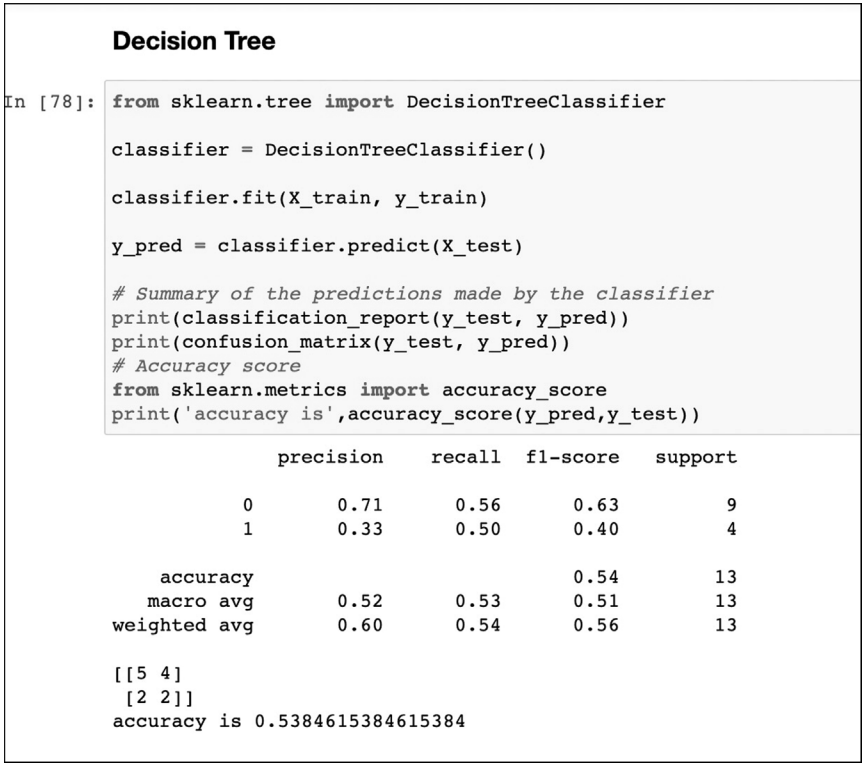

Fig. 13.27.    Decision Tree Model Result by Performance Metrics.

The choice of weights depends on the specific needs of the application. If the cost of a false positive is higher than the cost of a false negative, we may assign more weight to precision. Let's denote the weight for precision as $w\_p$ and the weight for recall as $w\_r$. Then, we can set the weight for $F_1$ score as 1, since it gives equal weight to both precision and recall. The weighted average can be expressed as:

$$\text{Weighted average} = \big(w\_p * \text{precision} + w\_r * \text{recall} + F_1 \text{ score}\big) \big/ \big(w\_p + w\_r + 1\big).$$

On the other hand, if the cost of a false negative is higher than the cost of a false positive, we may assign more weight to recall instead. In that case, we can set the weight for recall as $w\_r$ and the weight for precision as $w\_p = 1$. The weighted average becomes:

$$\text{Weighted average} = (\text{precision} + w\_r * \text{recall} + F_1 \text{ score})/(w\_r + 1).$$

We can get a more customised evaluation of the model's performance just by adapting the weights for each measure, which considers the specific needs of the application (Fig. 13.29).

**Bernoulli Naive Bayes**

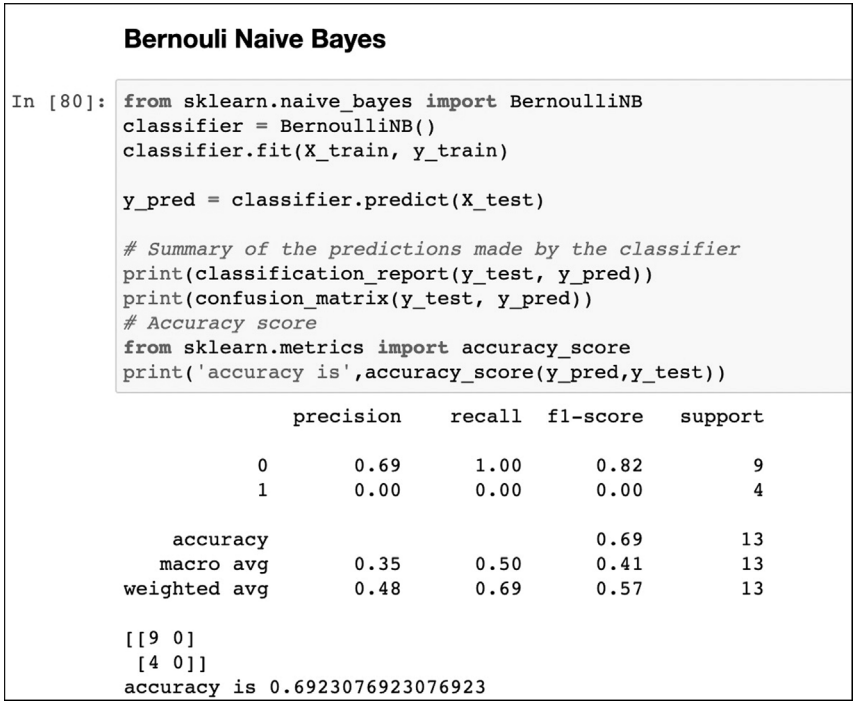

**Bernouli Naive Bayes**

```
In [80]: from sklearn.naive_bayes import BernoulliNB
         classifier = BernoulliNB()
         classifier.fit(X_train, y_train)

         y_pred = classifier.predict(X_test)

         # Summary of the predictions made by the classifier
         print(classification_report(y_test, y_pred))
         print(confusion_matrix(y_test, y_pred))
         # Accuracy score
         from sklearn.metrics import accuracy_score
         print('accuracy is',accuracy_score(y_pred,y_test))

                       precision    recall  f1-score   support

                    0       0.69      1.00      0.82         9
                    1       0.00      0.00      0.00         4

             accuracy                           0.69        13
            macro avg       0.35      0.50      0.41        13
         weighted avg       0.48      0.69      0.57        13

         [[9 0]
          [4 0]]
         accuracy is 0.6923076923076923
```

Fig. 13.28.    Bernoulli Naive Bayes Model Result by Performance
Metrics.

*Analysing the Model Built*

• Logistics regression has an accuracy of 046%, which is low and below average,
  hence the performance is considered below average or rather not a good model.
• Support vector machine (SVM) has an accuracy of 70%, which is above
  average, and the model performance is considered high or rather a good model.
• Decision tree has an accuracy of 53%, which is above average, and the model
  performance is considered average or rather a good model.
• Lastly, Naive Bayes has an accuracy of 70%, which is above average, and the
  model performance is considered high or rather a good model (Fig. 13.28).

## Model Evaluation

### 9. Model Evaluation

```
In [85]:  from sklearn.metrics import accuracy_score, log_loss
          classifiers = [
                    LogisticRegression(),
                    DecisionTreeClassifier(),
                    BernoulliNB(),
                    SVC(),
                            ]

          # Logging for Visual Comparison
          log_cols=["Classifier", "Accuracy", "Log Loss"]
          log = pd.DataFrame(columns=log_cols)

          for clf in classifiers:
              clf.fit(X_train, y_train)
              name = clf.__class__.__name__
              print("="*30)
              print(name)

              print('****Results****')
              train_predictions = clf.predict(X_test)
              acc = accuracy_score(y_test, train_predictions)
              print("Accuracy: {:.4%}".format(acc))

              log_entry = pd.DataFrame([[name, acc*100, 11]], columns=log_cols
              log = log.append(log_entry)

              print("="*30)

          ==============================
```

```
==============================
LogisticRegression
****Results****
Accuracy: 46.1538%
==============================
==============================
DecisionTreeClassifier
****Results****
Accuracy: 53.8462%
==============================
==============================
BernoulliNB
****Results****
Accuracy: 69.2308%
==============================
==============================
SVC
****Results****
Accuracy: 69.2308%
==============================
```

Fig. 13.29.    Model Evaluation.

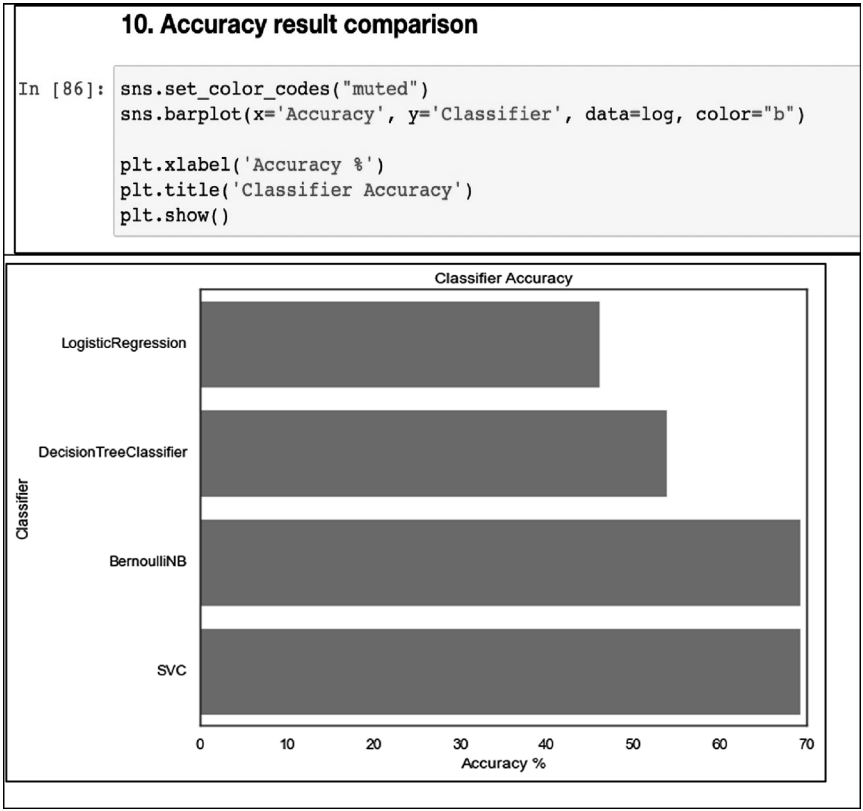

**10. Accuracy result comparison**

```
In [86]:   sns.set_color_codes("muted")
           sns.barplot(x='Accuracy', y='Classifier', data=log, color="b")

           plt.xlabel('Accuracy %')
           plt.title('Classifier Accuracy')
           plt.show()
```

Fig. 13.30.    Accuracy Comparison.

### Accuracy Result Comparison

The best models for categorising the governance of Edmonton are using SVC or Bernoulli Naive Bayes as they have a high performance of accuracy of 70%, while the rest of the models are average (Figs. 13.30 and 13.31).

### Exploratory Analysis

Exploratory analysis by making different types of dashboards in Power BI (Business intelligence) which is a method used to gain a preliminary understanding of a data set. It involves visualising the data and summarising its main features, such as the mean, median, and standard deviation, to identify patterns and anomalies.

A critical analysis in EDA involves going beyond the simple summary statistics and visualisations and examining the underlying structure of the data. This can involve identifying correlations and relationships between variables, discovering hidden patterns and trends, and testing assumptions about the data. It is a crucial step in the data analysis process as it helps to identify potential issues with the data, such as missing values or outliers, which can impact the results of any subsequent statistical analysis.

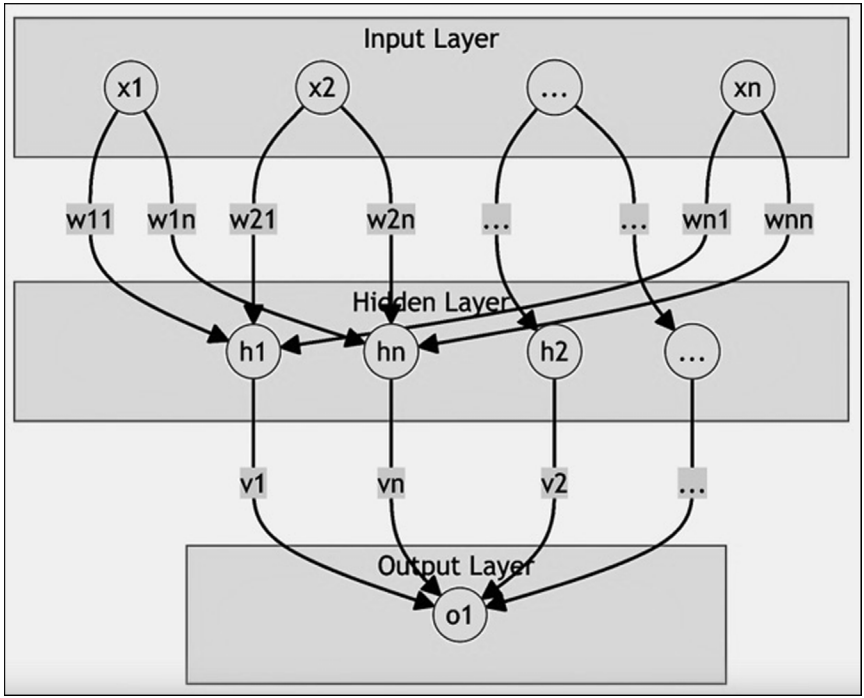

Fig. 13.31.    Artificial Neural Network (ANN).

**Heatmap**

The above heat map (Fig. 13.32) gives a glimpse of a strong correlation amongst the variables which vary in terms of co-efficient from 1 to −0.77. The unemployment rate is firmly correlated with other variables but impaired driving incidents and working-age population growth.

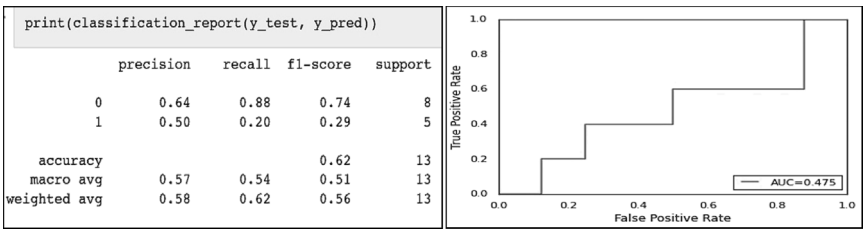

Fig. 13.32.    ANN confusion metrics accuracy result and ROC /AUC curve.

**Average**
Average of key factors by governance (see Fig. 13.33).

To assist leadership decision-making, the dashboard gives an average of important governance criteria, such as citizen participation, data protection, and regulatory compliance. The dashboard determines the average score of these factors for each governance category (such as local government, state government, and federal government), compares how well each governance category is performing, and pinpoints areas where policies and practises of the government could be improved (Fig. 13.29).

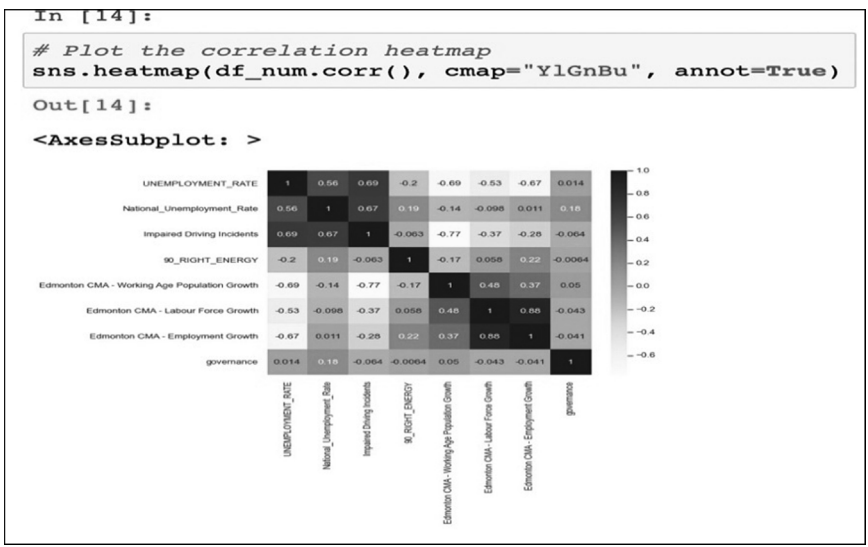

Fig. 13.33.  Correlation Analysis by Heatmap.

## Time Series Dashboard

Time series graph for various factors affecting the smart city (see Fig. 13.34).

Here, the dashboard displays recurring trends and patterns in the numerous variables influencing the smart city.

Decision-makers can determine if variables are becoming better or worse and whether changes are coming slowly or quickly by examining time series graphs. By giving decision-makers a visual depiction of the elements influencing the smart city, it facilitates the making of data-driven decisions. The dashboard shows the connections and causes between many elements that influence the smart city. The dashboard illustrates how to predict future trends and cyclical swings in the variables affecting things like traffic volume or energy usage in the smart city, allowing decision-makers to prepare ahead and take preventative measures to solve prospective difficulties.

## Prediction

Outliers are located using scatter plots, which are also used to predict values

Fig. 13.35 Forecasting on a specific major area which predominantly focuses on a smart city by using historical data (last 10 years). In power BI.

## Forecasting Unemployment

Forecasting the unemployment rate in a smart city can assist decision-makers in resource allocation, economic health assessment, social issue identification, investment stimulation and sustainable growth. This will ultimately secure a sustainable and inclusive smart city future for all members of society.

According to the above dashboard in the future unemployment rate will increase to 60.38 in 2032 and 54.40 in 2022 (see Fig. 13.36).

So, forecasting how will help society at present to take further action for making a smart city. And that decision a leader can only take at the right time if they have a digital smart city dashboard which will help them to make a sustainable future Forecasting unemployment rate in a smart city can also help to identify potential economic issues, target resources to areas in need, support sustainable growth, address social issues and encourage investment in the city.

### *Forecasting the Use of Solar Panel*

Forecasting the use of solar panel dashboards can assist citizens of smart cities in cutting back on their energy usage, and costs, adopting renewable energy sources, and lessening their reliance on fossil fuels, resulting in more efficient and economical energy consumption, all of which are significant outcomes for creating a sustainable environment. So, in order to save people's lives more conveniently, a leader uses the forecasting method to make a quick decision from a digital smart city dashboard (see Fig. 13.36).

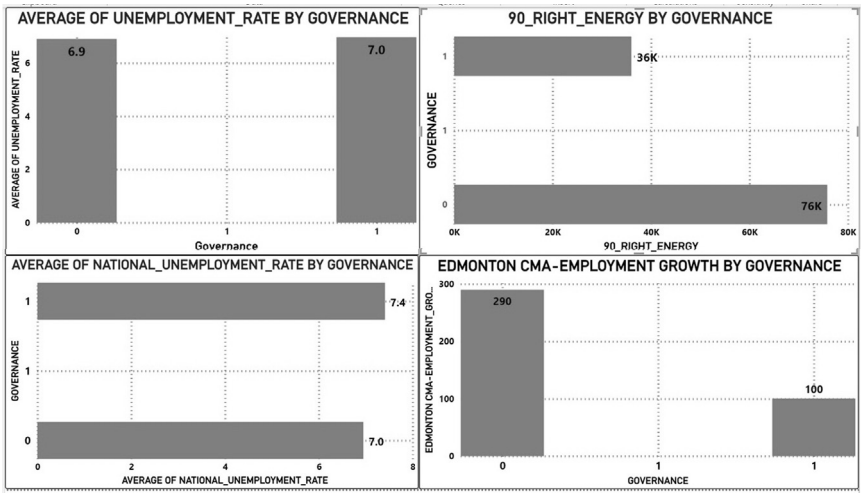

Fig. 13.34.   Average of Key Factors by Governance.

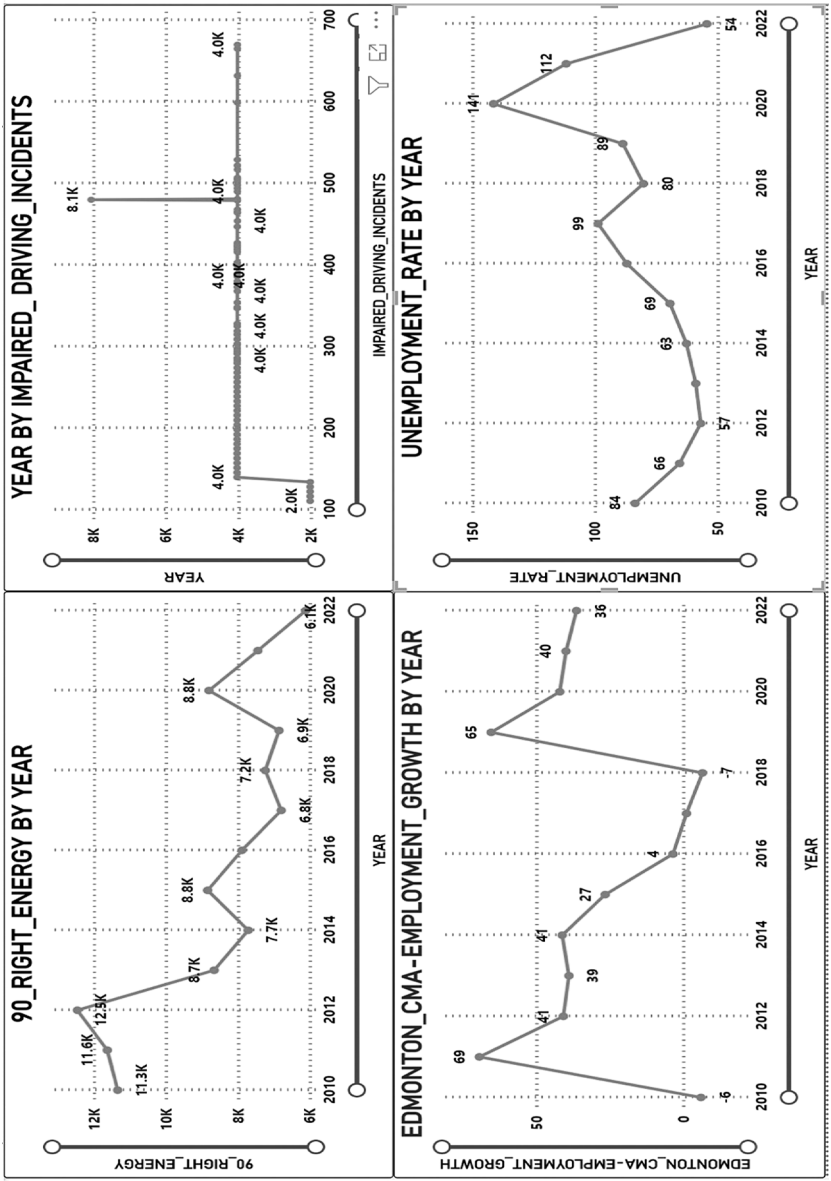

Fig. 13.35.    Time Series Graph.

The rate of energy users/resources, as seen above the smart city screen, was 679.29 in 2022 and 669.29 in 2032. Therefore, a leader can make a swift decision that results in the creation of a smart city where people can benefit from their living circumstances because of the relevance of the upper bound and lower bound range.

**Drawbacks**

The main drawback is that the database for Edmonton has several data sets with some having very few observations and hence not suitable for analysis or machine learning, while other data frames are not compatible.

*Limitations and Prospects*

The analysis was limited to five variables or attributes with no distinguished connection or rather an effect. While it is necessary to first establish a problem before attempting to solve it, this analysis had a problem statement but not a well-defined data set that would be used to provide a solution. This is because data sets in the database are not well categorised (Fig. 13.37).

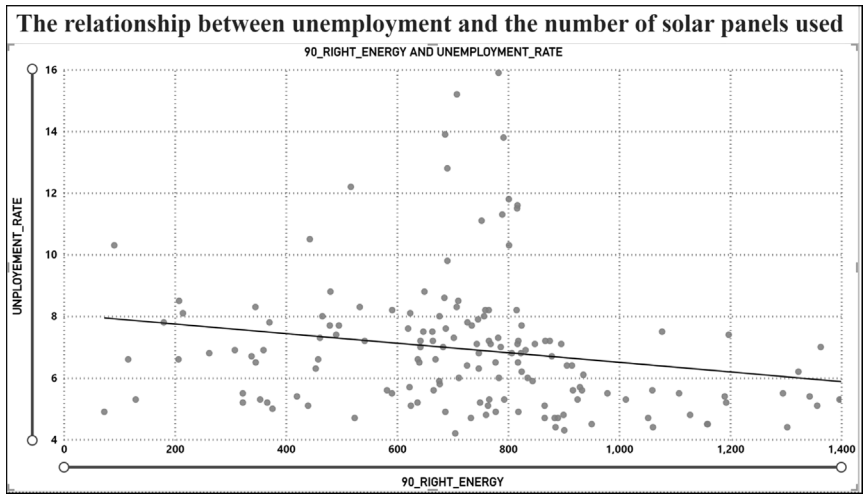

Fig. 13.36.    Relationship Between Unemployment and Number of Solar Panels Used.

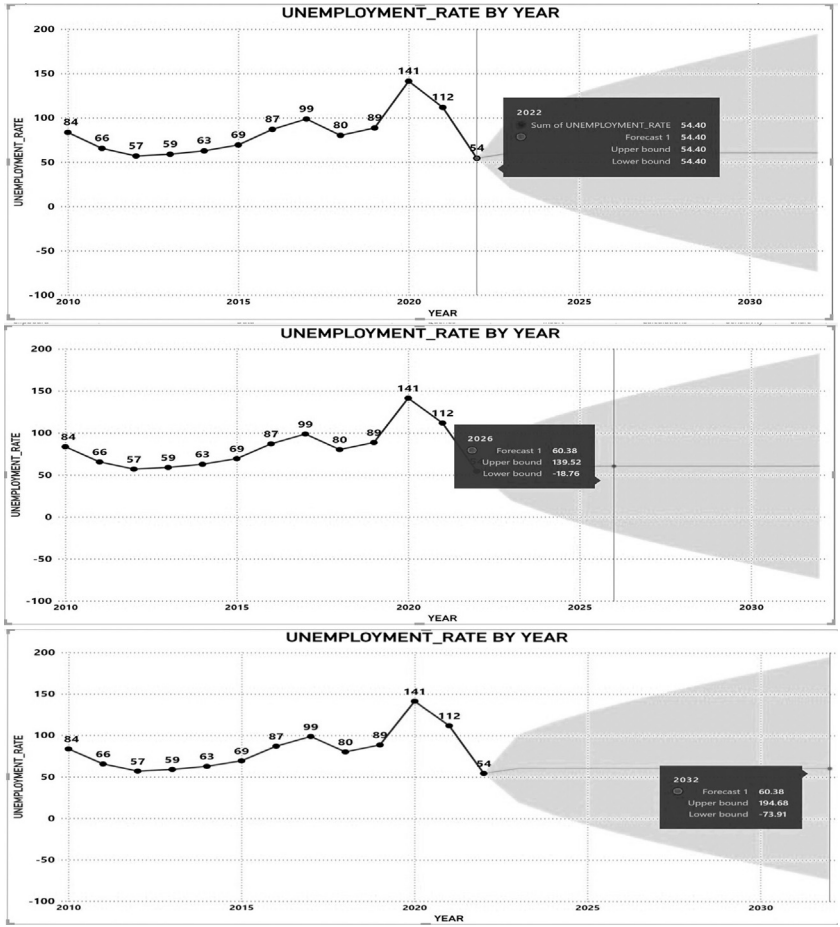

Fig. 13.37.    Unemployment Rate Trends Over Years.

### *Potential Impact on Community Development and City Dashboard*

The analysis highlights the significance of providing inhabitants with accurate data visualisations of the quality of life in various city neighbourhoods through a variety of categories that will have varying priorities for various user groups. This can improve the general quality of life in the city by encouraging residents to take an active role in decision-making regarding their surroundings. Municipalities gain from accurate, current data and tools for evaluating urban plans and projects, while other organisations and the private sector can increase the value of their offerings by leveraging the data (Fig. 13.38). Although many cities employ appealing and interesting visualisation techniques, the analysis of dashboards revealed that many of them frequently lack the tools necessary for choosing and analysing the data, creating their own indicators, or providing metadata or descriptions to aid in the interpretation or use of the data. The potential of geo-analytical and IT technologies for creating a genuine dialogue with residents is still largely untapped. The advantages of employing dashboards for bettering urban living can be increased by addressing these areas (Pluto-Kossakowska et al., 2022). In today's very diversified urban environment, the use of professional and well-informed decision-making tools is crucial for the management and governance of contemporary cities (Kourtit, 2018). The authors emphasise the application of an i-city dashboard based on big data and the tactical Pentagon model as a tool for managing intricate and unpredictable urban dynamics. Their empirical study, which used data from Stockholm, demonstrated the viability and value of the urban dashboard in helping decision-makers make well-informed judgements. To attain sustainability and peak performance in the urban system, the authors contend that additional research is required to create a fully developed, mature dashboard system and to create a solid link between policies, plans, and stakeholders. Cause-and-effect interactions should be given the attention they deserve to identify issues, spot possibilities, and evaluate the success and value creation in the urban system. Digital information technology should be created in unison with municipal demands and potential.

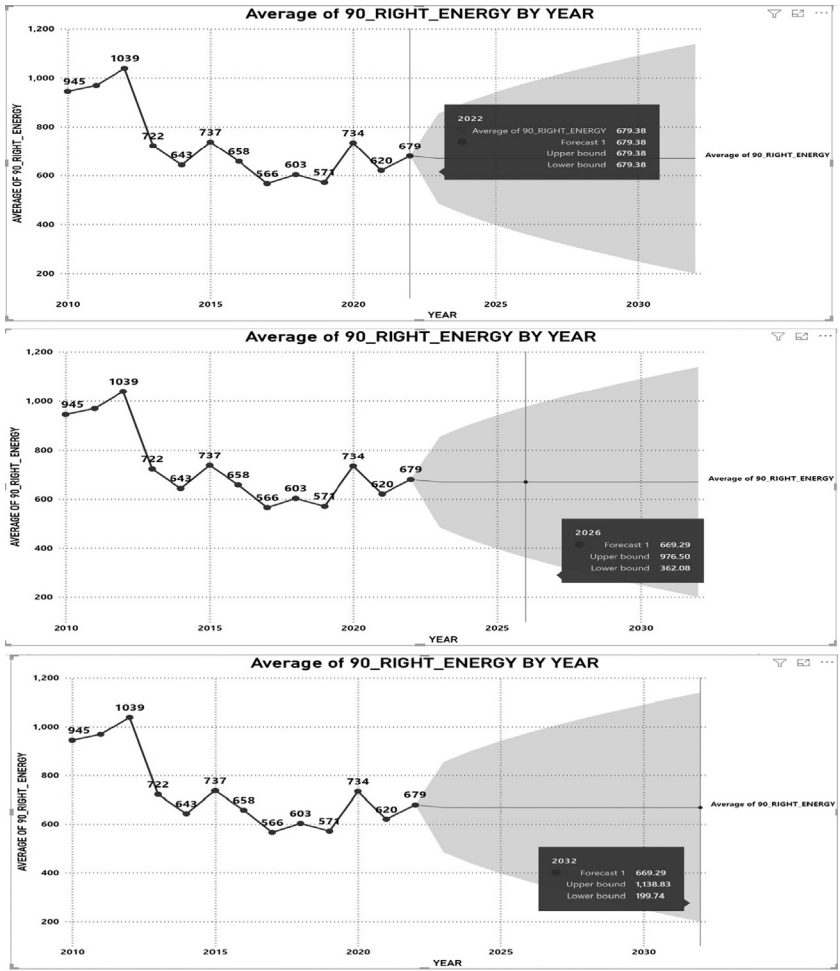

Fig. 13.38.    Average Right Energy Over Years.

## Conclusion

This study presents an innovative Smart City Performance Dashboard, designed to evaluate the impact of smart city initiatives on residents' quality of life. This tool leverages artificial intelligence and machine learning models such as Logistic Regression, Support Vector Machine, Decision Tree, Naive Bayes, and Artificial Neural Networks (ANN) for analyzing key performance indicators. These analyses assist city leaders in making strategic decisions and managing urban services more effectively. It serves as a robust tool for in-depth analysis of complex, dynamic urban systems, evaluating their success and generated value. Our tailored dashboard promotes investment in smart city initiatives by facilitating their evaluation. By analyzing complex urban systems and the influence of smart programs on daily life, we aim to connect stakeholders, policies, and strategies for sustainable urban development and performance optimization. The concept of 'smart cities,' central to our study, involves using cutting-edge technologies like Information and Communication Technology (ICT) to improve urban life. We believe a comprehensive governance framework is crucial for the successful implementation of smart city initiatives. Through this work, we offer a deeper understanding of city dashboards' role in governing smart cities. The demonstrated use of AI and machine learning models in this context is anticipated to inspire the development of innovative tools and systems for managing smart cities worldwide.

## Appendix

(1) Dataset:     https://github.com/AminRana/City-Dashboard-for-smart-city-governance

(2) Python   Code:   https://github.com/AminRana/City-Dashboard-for-smart-city-governance